\documentclass[a4paper,11pt]{article}
\usepackage[a4paper,left=2.73cm,right=2.7cm,top=3cm,bottom=3.5cm]{geometry}

\usepackage{amsfonts,amsmath,amssymb}
\usepackage{graphicx,caption,subcaption}
\usepackage[usenames, dvipsnames]{color}

\usepackage[colorlinks=true,linktocpage=true,linkcolor=blue,citecolor=blue]{hyperref}

\addtocontents{toc}{\protect\setcounter{tocdepth}{2}}
\numberwithin{equation}{section}

\def\cI{{\cal I}}

\def\cN{{\cal N}}
\def\cQ{{\cal Q}}
\def\cR{{\cal R}}

\newcommand{\mt}[1]{\textrm{\tiny #1}}

\newcommand{\be}{\begin{equation}}
\newcommand{\ee}{\end{equation}}

\begin{document}

\begin{titlepage}

\thispagestyle{empty}

\begin{center}

{\LARGE \textbf{BPS black holes from massive IIA on S$^6$}}

\vspace{30pt}
		
{\large \bf Adolfo Guarino  and Javier Tarr\'\i o}
		
\vspace{25pt}
		
{\normalsize  
Universit\'e Libre de Bruxelles (ULB) and International Solvay Institutes,\\
Service  de Physique Th\'eorique et Math\'ematique, \\
Campus de la Plaine, CP 231, B-1050, Brussels, Belgium.}

\vspace{20pt}

\texttt{aguarino} and \texttt{jtarrio}  \texttt{@ ulb.ac.be}

\vspace{20pt}
				
\abstract{
\noindent  We present BPS black hole solutions in a four-dimensional $\,\cN=2\,$ supergravity with an abelian dyonic gauging of the universal hypermultiplet moduli space. This supergravity arises as the $\,\textrm{SU}(3)$-invariant subsector in the reduction of massive IIA supergravity on a six-sphere. The solutions are supported by non-constant scalar, vector and tensor fields and interpolate between a unique $\,\textrm{AdS}_{2} \,\times\, \textrm{H}^2\,$ geometry in the near-horizon region and the domain-wall $\,\textrm{DW}_{4}\,$ (four-dimensional) description of the \mbox{D2-brane} at the boundary. Some special solutions with charged AdS$_{4}$ or non-relativistic scaling behaviours in the ultraviolet are also presented.
}

\end{center}

\end{titlepage}

\tableofcontents

\hrulefill
\vspace{10pt}

\section{Motivation and outlook}
\label{section:intro}

The search for BPS black hole solutions in four-dimensional $\,\mathcal{N}=2\,$ gauged supergravities with an embedding in string/M-theory  has recently captured new attention in light of the gravity/gauge correspondence. 

An interesting program started with the classification of asymptotically AdS$_{4}$ black holes in $\,\mathcal{N}=2\,$ supergravity coupled to vector multiplets in the presence of U(1) Fayet--Iliopoulos (FI) gaugings and non-constant scalar fields \cite{Cacciatori:2008ek,Cacciatori:2009iz}. The case with three vector multiplets \mbox{(STU model)}, a square root prepotential and all the FI parameters identified, corresponds to the U(1)$^4$-invariant subsector \cite{Duff:1999gh,Cvetic:1999xp} of the maximal \mbox{SO(8)-gauged} supergravity~\cite{deWit:1982ig}. 
This supergravity arises from the reduction of eleven-dimensional supergravity on a seven-sphere~\cite{deWit:1986iy}, and has a maximally supersymmetric AdS$_{4}$ solution dual to the three-dimensional ABJM superconformal field theory \cite{Aharony:2008ug} at low ($k=1,2$) Chern-Simons (CS) levels $\,k\,$ and $\,-k\,$. When uplifted to eleven dimensions, this solution corresponds to the Freund-Rubin $\,{\textrm{AdS}_{4} \times \textrm{S}^{7}}\,$ vacuum \cite{Freund:1980xh} describing the near-horizon geometry of the M2-brane.  
A charged version of this AdS$_{4}$ vacuum corresponds to the ultraviolet behaviour of the BPS black holes constructed in \cite{Cacciatori:2008ek,Cacciatori:2009iz} (\mbox{see refs~\cite{Halmagyi:2013sla,Erbin:2014hsa,Monten:2016tpu}} for M-theory models also containing hypermultiplets). In contrast, the infrared behaviour approaches an $\,\textrm{AdS}_{2} \times \textrm{S}^{2}\,$ geometry \cite{Cucu:2003yk} with the scalars determined by the attractor mechanism \cite{Cacciatori:2009iz,DallAgata:2010ejj,Hristov:2010ri}. The holographic interpretation is an RG flow across dimensions, more specifically, between a CFT$_{3}$ and a CFT$_{1}$. Using supersymmetric localisation techniques, a counting of microstates of BPS black holes in AdS$_4$ was performed in the dual field theory \cite{Benini:2015eyy,Benini:2016rke}\cite{Cabo-Bizet:2017jsl} -- identified as a deformation of the ABJM theory by a topological twist \cite{Benini:2015noa,Hosseini:2016tor,Hosseini:2016ume} -- and it was shown to match the Bekenstein--Hawking entropy~\cite{Bekenstein:1973ur,Hawking:1974sw}. 

This correspondence also has a realisation on the D3-brane of the type IIB theory, once the latter is reduced on a five-sphere to a five-dimensional maximal SO(6)-gauged supergravity \cite{Gunaydin:1985cu}. In this case, solutions interpolating between $\,\textrm{AdS}_{5}\,$ and $\,\textrm{AdS}_{3} \times \Sigma_{2}\,$ geometries, with $\,\Sigma_{2}\,$ being a Riemann surface, have a holographic interpretation in terms of RG flows between a CFT$_{4}$ and a CFT$_{2}$ \cite{Benini:2012cz,Benini:2013cda,Benini:2015bwz}\cite{Hosseini:2016cyf}. The field theory dual is a topologically twisted $\,\mathcal{N} = 4\,$ super Yang-Mills theory (SYM).

The present paper continues this program and classifies BPS black hole solutions in the $\,{\mathcal{N}=2}\,$ subsector of the four-dimensional maximal ISO(7)-gauged supergravity studied in~\cite{Guarino:2015qaa}. This supergravity arises in the reduction of the massive IIA theory on a six-sphere~\cite{Guarino:2015jca,Guarino:2015vca}. We focus on the SU(3)-invariant subsector which is described by an $\,\mathcal{N}=2\,$ supergravity coupled to a vector multiplet and the universal hypermultiplet \cite{Cecotti:1988qn} (see Table~\ref{table:N=2_model}). Because of the massive IIA origin, this setup differs from the \mbox{M-theory} and type IIB cases discussed before. For instance, the massive IIA theory has a $\,\textrm{DW}_{4}\,$ domain-wall solution (instead of an AdS$_{4}$ vacuum) as the four-dimensional description of the near-horizon limit of the \mbox{D2-brane}~\cite{Guarino:2016ynd}. Such a $\,\textrm{DW}_{4}\,$ solution is the non-conformal analog of the AdS$_{4}$ (AdS$_{5}$) vacuum in the M-theory (type IIB) models, and thus controls the ultraviolet behaviour of generic BPS flows.

In this paper we present a two-parameter family of BPS black hole solutions that feature a unique $\,\textrm{AdS}_{2} \times \textrm{H}^{2}\,$ geometry in the infrared and flow to a charged version of the $\,\textrm{DW}_{4}\,$ solution describing the D2-brane in the ultraviolet. The scalar fields in the vector multiplet and hypermultiplet are non-constant along the flow and enter the black hole horizon as dictated by the attractor equations. For specific values of the parameters, the solutions flow to either an $\,\mathcal{N}=2\,$ charged AdS$_{4}$ vacuum or a non-relativistic metric in the ultraviolet \cite{Halmagyi:2011xh,Chimento:2015rra,Cardoso:2015wcf}, instead of to the generic charged $\,\textrm{DW}_{4}\,$ solution. It would be very interesting to understand these flows from a dual field theory perspective using the massive IIA on S$^{6}$/SYM-CS duality~\cite{Guarino:2015jca,Schwarz:2004yj}.

\section{$\mathcal{N}=2$ supergravity with abelian gaugings from massive IIA}
\label{section:N=2model}

Massive IIA ten-dimensional supergravity admits a consistent truncation on the six-sphere \cite{Guarino:2015vca} to maximal $D=4$ supergravity with a dyonic ISO(7) gauging \cite{Guarino:2015qaa}. Within this truncation, there is a subsector that is invariant under the action of an $\textrm{SU}(3)$ subgroup of the $\textrm{ISO}(7)$ gauge group, and is given by an $\,\cN=2\,$ supergravity coupled to a vector multiplet and the universal hypermultiplet \cite{Guarino:2015qaa}. The dynamical (bosonic) degrees of freedom of this $\,\cN=2\,$ subsector are summarised in Table~\ref{table:N=2_model}.
\begin{table}[t!]
\centering
\scalebox{1}{
\begin{tabular}{cccccc}
\hline
spin & gravity multiplet & vector multiplet & universal hypermultiplet \\
\hline
$2$ & $g_{\mu\nu}$  &   &  \\[1pt]
$1$ & $\mathcal{A}_{\mu}^0$  & $\mathcal{A}_{\mu}^1$  &  \\[1pt]
$0$ &   & $\chi$ \,,\, $\varphi$  & $\phi$ \,,\, $a$ \,,\, $\zeta$ \,,\, $\tilde \zeta$ \\
\hline
\end{tabular}}
\caption{Bosonic fields in the $\cN=2$ and SU(3)-invariant subsector of the maximal supergravity multiplet in four dimensions. 
}
\label{table:N=2_model}
\end{table}

We follow closely  the $\,\mathcal{N}=2\,$ supergravity conventions of \cite{Klemm:2016wng} except for a change of gauge in the ansatz for the vector and tensor fields, to be discussed below. The two real scalars in the vector multiplet (see Table~\ref{table:N=2_model}) can be grouped into a complex one
\be
z \equiv - \chi + i e^{-\varphi} \ ,
\ee
describing the special K\"ahler manifold $\,\mathcal{M}_{\textrm{SK}}=\textrm{SU(1,1)}/\textrm{U}(1)\,$ in terms of holomorphic sections $\,{X^{M}(z)=(X^{\Lambda}(z),F_{\Lambda}(z))}\,$. Here $\,M\,$ is a symplectic Sp(4) vector index, whereas $\,\Lambda=0,1\,$  runs over the first (electric) or second (magnetic) half of components. 
It  proves convenient to define a symplectic product of vectors  
\be
\left\langle U , V \right\rangle \equiv  U^{M} \Omega_{MN}V^{N} = U_{\Lambda} V^{\Lambda} - U^{\Lambda} V_{\Lambda} \ ,
\ee
where $\,\Omega_{MN}\,$ is the antisymmetric invariant matrix of Sp(4). In terms of it, the K\"ahler potential associated to $\,\mathcal{M}_{\textrm{SK}}\,$ can be expressed as $\,K=-\log(i \left\langle X, \bar{X} \right\rangle)\,$.
In the $\,\mathcal{N}=2\,$ model studied in \cite{Guarino:2015qaa} the sections take the form
\be
{(X^{0}\,,\,X^{1}\,,\,F_{0}\,,\,F_{1})} =  {(-z^3\,,\,-z\,,\,1\,,\,3z^2)} \ , 
\ee
and satisfy the relation $\,F_{\Lambda}=\partial \mathcal{F}/\partial X^{\Lambda}\,$ for a prepotential $\,\mathcal{F}\,$ of the form
\begin{equation}
\label{F_prepotential}
\mathcal{F}=-2\sqrt{X^{0}(X^{1})^3} \ ,
\end{equation}
whereas the K\"ahler potential yields a K\"ahler metric of the form 
\be
ds^2_{\textrm{SK}} = - K_{z\bar{z}} \, d z \, d\bar{z}   = - \frac{3}{4} \frac{d z \, d\bar{z}}{(\textrm{Im} z)^2}  \ .
\ee

The generalised theta angles and coupling constants for the vector fields entering the Lagrangian are encoded in a complex matrix that depends only on the scalar $z$
\begin{equation}
\label{N-matrix}
\mathcal{N}_{\Lambda \Sigma} = \bar{F}_{\Lambda \Sigma}+ 2 i \frac{\textrm{Im}(F_{\Lambda \Gamma})X^{\Gamma}\,\,\textrm{Im}(F_{\Sigma \Delta})X^{\Delta}}{\textrm{Im}(F_{\Omega \Phi})X^{\Omega}X^{\Phi}}
\qquad \textrm{ with } \qquad
F_{\Lambda \Sigma}=\partial_{\Lambda}\partial_{\Sigma} \mathcal{F} \ .
\end{equation}
Extracting $\,\mathcal{R}_{\Lambda \Sigma}\equiv \textrm{Re}(\mathcal{N}_{\Lambda \Sigma})\,$ and $\,\mathcal{I}_{\Lambda \Sigma}\equiv \textrm{Im}(\mathcal{N}_{\Lambda \Sigma})\,$ from (\ref{N-matrix}), we introduce a scalar matrix $\mathcal{M}_{MN}(z)$ that restores symplectic covariance and will be relevant later on when presenting the BPS equations. It takes the form
\begin{equation}
\label{M-matrix}
\mathcal{M}(z) = \left( 
\begin{array}{cc}
\mathcal{I} + \mathcal{R} \mathcal{I}^{-1} \mathcal{R}   & -\mathcal{R} \mathcal{I}^{-1} \\
- \mathcal{I}^{-1} \mathcal{R} & \mathcal{I}^{-1}
\end{array}
\right) \ ,
\end{equation}
and satisfies $\,\mathcal{M}_{MN} \mathcal{V}^{N}=i \Omega_{MN}\mathcal{V}^{N}\,$ and $\,\mathcal{M}_{MN} D_{z}\mathcal{V}^{N}=-i \Omega_{MN}D_{z}\mathcal{V}^{N}$, where $\,\mathcal{V}^{M} \equiv e^{K/2} \, X^{M}\,$ is a redefined (non-holomorphic) set of symplectic sections with K\"ahler covariant derivatives given by $\,D_{z}\mathcal{V}^{M} = \partial_{z} \mathcal{V}^{M} + \frac{1}{2} (\partial_{z} K) \mathcal{V}^{M}\,$.

Consider now the universal hypermultiplet $\mathcal{M}_{\textrm{QK}}=\textrm{SU(2,1)}/(\textrm{SU}(2) \times \textrm{U}(1))$. The four real scalars spanning this quaternionic K\"ahler geometry are collectively denoted $q^{u}=(\phi,a,\zeta,\tilde{\zeta})$, with metric
\begin{equation}
ds_{\textrm{QK}}^2 = \, - h_{uv} \, dq^{u} dq^{v} 
= -  d\phi^2  - \frac{1}{4} e^{4 \phi} \left( d a + \frac{1}{2} \left( \zeta \, d \tilde \zeta - \tilde \zeta \, d \zeta \right) \right)^2   - \frac{1}{4} \, e^{2\phi} \left( d \zeta^2 + d \tilde \zeta^2 \right)  \ .
\end{equation}
The specific $\,\mathcal{N}=2\,$ models that we focus on involve an abelian $\,\mathbb{R} \times \textrm{U(1)}\,$ gauging of two isometries of this quaternionic manifold. The relevant Killing vectors $k_{\alpha}$ (where $\alpha=\mathbb{R}$ or $\textrm{U}(1)$) are
\begin{equation}
\label{Killing-vectors}
k_{\small \mathbb{R}} = \partial_{a} \ , \qquad
k_{\small \textrm{U}(1)} = 3 (\zeta \partial_{\tilde{\zeta}} - \tilde{\zeta} \partial_{\zeta}) \ ,
\end{equation}
and can be derived from an SU(2) triplet of moment maps $\,\mathcal{P}^{x}_{\alpha}\,$ of the form
\begin{equation}
\mathcal{P}^{x}_{\mathbb{R}} = ( \, 0 \,,\, 0 \,,\, - \tfrac{1}{2} e^{2\phi} \,) \ , \qquad
\mathcal{P}^{x}_{\textrm{U}(1)} = 3 \, \Big( \, - e^{\phi} \tilde{\zeta} \, , \, e^{\phi} \zeta \,,\,  1-\tfrac{1}{4} e^{2\phi} (\zeta^2+\tilde{\zeta}^2) \, \Big) \ .
\end{equation}

The gaugings under consideration in this work are of the dyonic type first introduced in \cite{Dall'Agata:2012bb} and further explored in \cite{Dall'Agata:2014ita}. These gaugings involve both  electric $\,\mathcal{A}_{\mu}{}^{\Lambda}\,$ and magnetic $\,{\tilde{\mathcal{A}}}_{\mu \, \Lambda} \,$ vector fields as gauge connections in the covariant derivatives. The vector fields can be arranged into an Sp(4) symplectic vector ${\mathcal{A}_{\mu}{}^{M}= ( \mathcal{A}_{\mu}{}^{\Lambda} , {\tilde{\mathcal{A}}}_{\mu \, \Lambda} ) } $ in terms of which the covariant derivatives for the scalars in the hypermultiplet read
\begin{equation}
\label{Dq}
D_{\mu} q^{u} = \partial_{\mu} q^{u} - \mathcal{A}_{\mu}{}^{M}  \, \Theta_{M}{}^{\alpha} \, k_{\alpha}{}^{u} = \partial_{\mu} q^{u} - \mathcal{A}_{\mu}{}^{M}  \, \mathcal{K}_{M}{}^{u} \ .
\end{equation}
Following \cite{Klemm:2016wng}, we have introduced Killing vectors of the form $\,\mathcal{K}_{M}{}^{u}  \equiv \Theta_{M}{}^{\alpha} \, k_{\alpha}{}^{u}\,$ in (\ref{Dq}) in order to restore symplectic covariance.

The embedding tensor $\,\Theta_{M}{}^{\alpha}\,$ in (\ref{Dq}) is constant and specifies the linear combinations of electric and magnetic vectors that enter the gauge connection. Consistency requires a quadratic constraint on the embedding tensor of the form $\,\left\langle\Theta^{\alpha} , \Theta^{\beta} \right\rangle=0\,$ \cite{deWit:2005ub}. This constraint can be viewed as an orthogonality condition between the charges $\,\Theta_{M}{}^{\alpha}\,$ in \eqref{Dq}, and guarantees that a dyonic gauging involving electric and magnetic vectors can always be rotated back to a purely electric one by a change of symplectic frame. This change of symplectic frame is usually assumed in the literature in order to have a description involving electric vectors solely. However, a formulation in terms of a prepotential $\,\mathcal{F}\,$ might be no longer available after changing the symplectic frame. In this work, we  stay with the prepotential in (\ref{F_prepotential}) and do not perform any symplectic rotation to an electric frame. As a result, we deal with dyonic gaugings involving non-zero magnetic charges $\,\Theta^{\Lambda \alpha}\,$.

Consistency of the gauge algebra in the presence of magnetic charges requires one to introduce auxiliary two-form tensor fields $\,\mathcal{B}_{\mu\nu \, \alpha}\,$ that modify the field strengths of the dynamical vectors. For abelian gaugings, the latter are given by \cite{deWit:2005ub}
\begin{equation}
\label{H_def}
\mathcal{H}_{\mu\nu}{}^{\Lambda}=2 \, \partial_{[\mu} \mathcal{A}_{\nu]}{}^{\Lambda} - \frac{1}{2} \, \Theta^{\Lambda \alpha}\, \mathcal{B}_{\mu\nu \, \alpha} \ .
\end{equation}
Lastly, the tensor fields come along with their own set of tensor gauge transformations, which are intertwined with the ordinary vector gauge transformations. We will discuss the gauge fixing of this symmetry in the next section.

Using differential form notation, the bosonic Lagrangian that describes the dynamics of the dyonic gaugings of  $\mathcal{N}=2$ supergravity reads \cite{deWit:2005ub}
\begin{equation}
\label{Lagrangian_N2}
\begin{split}
L_{\cN=2} 	& =   \left( \frac{R}{2} - V_{g} \right) * \! 1 - K_{z\bar{z}} \, dz \wedge *  d\bar{z} - h_{uv} \, Dq^{u} \wedge *  Dq^{v}   \\[2mm]
& \quad + \frac{1}{2} \, \cI_{\Lambda \Sigma} \, \mathcal{H}^\Lambda \wedge \! * \mathcal{H}^\Sigma +  \frac{1}{2} \, \cR_{\Lambda \Sigma} \, \mathcal{H}^\Lambda \wedge \mathcal{H}^\Sigma \\[2mm]
& \quad + \frac{1}{2} \Theta^{\Lambda \alpha}\,  \mathcal{B}_{\alpha} \wedge d \tilde{\mathcal{A}}_{\Lambda} + \frac{1}{8} \, \Theta^{\Lambda \alpha} \, \Theta_{\Lambda}{}^{\beta} \, \mathcal{B}_{\alpha} \wedge \mathcal{B}_{\beta} \ ,
\end{split}
\end{equation}
where the last line is a topological term that is non-zero whenever magnetic charges $\,\Theta^{\Lambda \alpha}\,$ are present.\footnote{The expressions (\ref{H_def}) and (\ref{Lagrangian_N2}) match the ones given in \cite{deWit:2005ub} upon the identification $\,\mathcal{B}_{\alpha \, \tiny \textrm{[here]}} = -\mathcal{B}_{\alpha \, \tiny \cite{deWit:2005ub}}\,$. This is a consequence of the different convention adopted in \cite{Klemm:2016wng} and \cite{deWit:2005ub} for the antisymmetric matrix $\,\Omega_{MN}\,$.} Together with the Einstein-Hilbert term, and due to the abelian gauging in the hypermultiplet sector, the Lagrangian also contains a scalar potential $\,V_{g}\,$ given by
\begin{equation}
\label{Vg}
V_{g} =  4 \,  \mathcal{V}^{M}  \, \bar{\mathcal{V}}^{N}    \, \mathcal{K}_{M}{}^{u}  \, h_{uv} \,  \mathcal{K}_{N}{}^{v}
+ \mathcal{P}^{x}_{M} \, \mathcal{P}^{x}_{N} \left( K^{z\bar{z}} \, D_{z}\mathcal{V}^{M} \, D_{\bar{z}} \bar{\mathcal{V}}^{N}  - 3 \, \mathcal{V}^{M} \, \bar{\mathcal{V}}^{N} \right) \ ,
\end{equation}
where, as for the Killing vectors entering (\ref{Dq}), we have now introduced a symplectic vector of momentum maps $\,\mathcal{P}_{M}^{x}  \equiv \Theta_{M}{}^{\alpha} \, \mathcal{P}_{\alpha}^{x}\,$ in order to restore symplectic covariance \cite{Klemm:2016wng}. Therefore, the Lagrangian (\ref{Lagrangian_N2}) becomes completely specified in terms of the geometric data for $\,\mathcal{M}_{\textrm{SK}}\,$ and $\,\mathcal{M}_{\textrm{QK}}\,$ presented previously (Killing vectors, etc.), as well as a constant embedding tensor $\,\Theta_{M}{}^{\alpha}\,$ encoding the gauging of the theory.

\subsubsection*{The model of \cite{Guarino:2015qaa}}

The $\,\mathcal{N}=2\,$ dyonically gauged supergravity we explore in this work appears from the reduction of massive IIA supergravity on the six-sphere \cite{Guarino:2015qaa,Guarino:2015vca}. The corresponding gauging is determined by an embedding tensor $\,\Theta_{M}{}^{\alpha}\,$ of the form
\begin{equation}
\label{Theta_model}
\Theta_{M}{}^{\alpha} = \left(
\begin{array}{c}
\Theta_{\Lambda}{}^{\alpha} \\[2mm]
\hline\\[-2mm]
\Theta^{\Lambda \, \alpha} 
\end{array}\right)
= \left(
\begin{array}{cc}
\Theta_{0}{}^{\mathbb{R}} & \Theta_{0}{}^{\textrm{U(1)}} \\[2mm]
\Theta_{1}{}^{\mathbb{R}} & \Theta_{1}{}^{\textrm{U(1)}} \\[2mm]
\hline\\[-2mm]
\Theta^{0 \, \mathbb{R}} & \Theta^{0 \, \textrm{U(1)}} \\[2mm]
\Theta^{1 \, \mathbb{R}} & \Theta^{1 \, \textrm{U(1)}}
\end{array}\right)
= \left(
\begin{array}{cc}
g & 0 \\[2mm]
0 & g \\[2mm]
\hline\\[-2mm]
-m & 0 \\[2mm]
0 & 0
\end{array}\right) \ ,
\end{equation}
where $\,g\,$ and $\,m\,$ are constant parameters identified with the inverse radius of the six-sphere and with the Romans mass parameter, respectively, and are assumed to be positive. The parameter $\,g\,$ sources the electric part of the embedding tensor whereas the parameter $\,m\,$ activates the magnetic one. By setting $\,m=0\,$, the gauging is of electric type and the resulting $\mathcal{N}=2$ supergravity model has an uplift to the massless IIA theory (and thus also to M-theory). 

From the explicit form of the embedding tensor in \eqref{Theta_model} it follows that the $\,\mathbb{R}\,$ factor in the gauge group $\,\mathbb{R} \times \textrm{U}(1)\,$ is gauged dyonically by the vectors $\,{\mathcal{A}^0}\,$ and $\,\tilde{\mathcal{A}}_0\,$, whereas the U(1) factor is gauged only electrically by the vector $\,\mathcal{A}^{1}\,$. This can be seen from the covariant derivatives (\ref{Dq}) of the scalars in the universal hypermultiplet which, for our specific model, take the form
\begin{equation}
\label{Dq_model}
D a = d a + g \, \mathcal{A}^0 - m \, \tilde{\mathcal{A}}_0  \ ,  \qquad
D \zeta = d \zeta - 3 \, g \, \mathcal{A}^1 \tilde \zeta \ ,  \qquad
D \tilde \zeta = d \tilde \zeta + 3 \, g \, \mathcal{A}^1 \zeta \ .
\end{equation}
As a result, the shift symmetry associated with the Killing vector $\,k_{\small \mathbb{R}}=\partial_{a}\,$ in (\ref{Killing-vectors}) is gauged with the linear combination $\,\alpha^{-} \equiv g \, \mathcal{A}^0 - m \, \tilde{\mathcal{A}}_0\,$ of the graviphoton and its magnetic dual, whereas that of the $\,k_{\mathrm{U(1)}}\,$ Killing vector is gauged using the vector $\,\mathcal{A}^{1}\,$ in the vector multiplet, and the scalars $\,\zeta\,$ and $\,\tilde{\zeta}\,$ are charged under it. The model also contains a tensor field that modifies the electric field strengths according to (\ref{H_def}), resulting in
\begin{equation}
\mathcal{H}^0 = d \mathcal{A}^0 + \tfrac{1}{2} \,  m \, \mathcal{B}^0
\ , \qquad
\mathcal{H}^1 = d \mathcal{A}^1 \ ,
\end{equation}
where we have relabelled the tensor field as $\,\mathcal{B}^{0}\equiv \mathcal{B}_{\mathbb{R}}\,$. Therefore, the scalar $\,a\,$ in (\ref{Dq_model}) is a St\"uckelberg field, and the tensor field $\,\mathcal{B}^{0}\,$ becomes massive. Since the U(1) factor of the gauge group is gauged electrically only, the tensor field $\,\mathcal{B}_{\textrm{U}(1)}\,$ decouples from the system and can be consistently set to zero.

When particularised to the embedding tensor in (\ref{Theta_model}), the generic $\,\mathcal{N}=2\,$ supergravity Lagrangian in \eqref{Lagrangian_N2} becomes
\be\begin{split}
\label{Lagrangian_model}
L	& = \left( \frac{R}{2} - V_{g} \right) * \! 1 - \frac{3}{4} \left[ d \varphi \wedge \! * d \varphi + e^{2\varphi} \, d \chi \wedge \! * d  \chi \right] -  \, d \phi \wedge \! * d \phi \\
	& \quad - \frac{1}{4} e^{4 \phi} \left[ D a + \frac{1}{2} \left( \zeta \, D \tilde \zeta - \tilde \zeta \, D \zeta \right) \right] \wedge \! * \! \left[ D a + \frac{1}{2} \left( \zeta \, D \tilde \zeta - \tilde \zeta \, D \zeta \right) \right] \\
	& \quad - \frac{1}{4} \, e^{2\phi} \left[ D \zeta \wedge \! * D \zeta + D \tilde \zeta \wedge \! * D \tilde \zeta \right] + \frac{1}{2} \, \cI_{\Lambda \Sigma} \, \mathcal{H}^\Lambda \wedge \! * \mathcal{H}^\Sigma \\
	& \quad + \frac{1}{2} \, \cR_{\Lambda \Sigma} \, \mathcal{H}^\Lambda \wedge \mathcal{H}^\Sigma - \frac{1}{2} m\,  \mathcal{B}^{0} \wedge d \tilde{\mathcal{A}}_0 - \frac{1}{8} \, g \,  m\,  \mathcal{B}^{0} \wedge \mathcal{B}^{0} \ .
\end{split}\ee
It is important to note that the dyonic nature of the gauging implies the introduction of the magnetic vector $\,\tilde{\mathcal{A}}_0\,$ and the tensor field $\,\mathcal{B}^{0}\,$ which, however, does not affect the counting of degrees of freedom. These fields do not carry independent dynamics, as can be seen from the variations of the Lagrangian (\ref{Lagrangian_model}) with respect to them, which produce two first-order differential relations
\begin{equation}
\label{EOM_auxiliary}
\begin{split}
d \mathcal{B}^0  & =  - e^{4 \phi} * \! \left[ D a + \frac{1}{2} \left( \zeta \, D \tilde \zeta - \tilde \zeta \, D \zeta \right) \right] \ , \\
 d \tilde{\mathcal{A}}_0 + \frac{1}{2} \, g \,  \mathcal{B}^0 & =  \cI_{0\Lambda} * \! \mathcal{H}^\Lambda +  \cR_{0\Lambda} \, \mathcal{H}^\Lambda \ .
 \end{split}
\end{equation}
The former is a duality relation between the tensor field and the scalars in the universal hypermultiplet, whereas the later is the duality relation between the graviphoton and its magnetic dual. As anticipated below (\ref{H_def}), the introduction of the tensor field  comes along with an additional tensor gauge symmetry given by a one-form gauge parameter $\,\Xi^{0}\,$. Up to a total derivative, the Lagrangian (\ref{Lagrangian_model}) is invariant under the tensor gauge transformation
\begin{equation}
\label{tensor-gauge-transf}
\mathcal{B}^0 \to \mathcal{B}^0 - d  \Xi^{0} \ , \qquad
\mathcal{A}^0 \to \mathcal{A}^0 + \tfrac{1}{2} \, m \, \Xi^{0} \ , \qquad
\tilde{\mathcal{A}}_0 \to \tilde{\mathcal{A}}_0 + \tfrac{1}{2} \, g \, \Xi^{0} \ .
\end{equation}

Finally, plugging the embedding tensor (\ref{Theta_model}) into the expression of the scalar potential in (\ref{Vg}), and making again use of the scalar geometry data, one obtains
\be\begin{split}
\label{Vg_model}
V_{g} 	& = \frac{1}{8}\, g^2 \left[ e^{4 \phi - 3 \varphi} \left( 1 + e^{2 \varphi} \chi^2 \right)^3 - 12 \, e^{2 \phi - \varphi} \left( 1 + e^{2 \varphi} \chi^2 \right) - 24 \, e^\varphi  \right. \\
	& \qquad \qquad   + \frac{3}{4} \, e^{4 \phi + \varphi} \left( \zeta^2 + \tilde \zeta^2 \right)^2 \left( 1 + 3\, e^{2 \varphi} \chi^2 \right) + 3 \, e^{4 \phi + \varphi} \left( \zeta^2 + \tilde \zeta^2 \right) \chi^2 \left( 1 + e^{2 \varphi} \chi^2 \right) \\
	& \qquad \qquad \left. - 3 \, e^{2 \phi + \varphi} \left( \zeta^2 + \tilde \zeta^2 \right) \left( 1 - 3 \, e^{2 \varphi} \chi^2 \right) \right]  \\
	& \quad - \frac{1}{8} \, g \, m \, \chi \, e^{4 \phi + 3 \varphi} \left[ 3 \left( \zeta^2 + \tilde \zeta^2 \right) + 2 \, \chi^2 \right] + \frac{1}{8} \, m^2 \, e^{4 \phi + 3 \varphi} \ .
\end{split}\ee
The full set of equations of motion that follows from the $\,\mathcal{N}=2\,$ supergravity Lagrangian (\ref{Lagrangian_model}) is presented in  appendix~\ref{App:EOMs}.

\section{BPS equations in dyonically gauged $\,\mathcal{N}=2\,$ supergravity}

The generic Lagrangian (\ref{Lagrangian_N2}) of dyonically gauged $\,\mathcal{N}=2\,$ supergravity has recently been considered in \cite{Klemm:2016wng} to study static BPS flow equations with spherical $\textrm{S}^{2}$ ($\kappa=1$) or hyperbolic $\textrm{H}^{2}$ ($\kappa=-1$) symmetry. In this section we make extensive use of the results derived therein, and simply fetch the main results and equations needed to find BPS solutions in our model.

\subsection{Field ansatz and gauge fixing}
\label{sec:ansatz}

The most general metric compatible with sphericity/hyperbolicity and staticity is given by
\begin{equation}
\label{metric_general}
d s^2 = - e^{2 U(r)} d t^2 + e^{-2 U(r)} d r^2 + e^{2 (\psi(r) - U(r))} \left( d \theta^2 + \left( \frac{ \sin \sqrt{\kappa} \, \theta }{ \sqrt{ \kappa } } \right)^2 \, d \phi^2 \right) \ ,
\end{equation}
where we have partially-fixed diffeomorphisms by imposing that the radial component of the metric is the inverse of the temporal one. The functions $\,U(r)\,$ and $\,\psi(r)\,$ are assumed to depend solely on the radial coordinate $\,r\,$, and the same holds for the scalar fields $\,z(r)\,$ and $\,q^{u}(r)\,$. As we  show below (see eq.~(\ref{N=2_point})), the existence of a regular horizon in the infrared (IR)  imposes that the scalars $\,\zeta\,$ and $\,\tilde \zeta\,$ must vanish there. Furthermore, we will impose boundary conditions in the ultraviolet (UV) such that $\,\zeta\,$ and $\,\tilde \zeta\,$ vanish at $\,r \rightarrow \infty\,$. Then, by looking at the equations of motion in \eqref{eq.eomzetazetat} and at the form of $\,V_{g}\,$ in \eqref{Vg_model}, it is consistent to take
\begin{equation}
\label{zeta=barzeta=0}
\zeta(r) = \tilde \zeta(r) = 0 \ .
\end{equation}
From now on we restrict our study to configurations where this relation is imposed, which allows us to simplify the forthcoming discussion. This restriction also implies an enhancement of the residual symmetry of the SU(3)-invariant subsector of maximal supergravity to an $\,{\textrm{SU}(3)\times\textrm{U}(1)}\,$ symmetry as a consequence of turning off the scalar fields charged under the U(1) factor of the gauge group (see eq.~\eqref{Dq_model}).

Let us consider now the ansatz for the vector and tensor fields. For the vectors, staticity and spherical/hyperbolic symmetry of the associated field strengths imply that
\be
\label{eq.gaugeansatz}
\mathcal{A}^\Lambda = \mathcal{A}_t{}^\Lambda(r) \, d t - p^\Lambda \, \frac{ \cos \sqrt{\kappa} \, \theta }{ \kappa } \, d \phi \ ,
\ee
with $\,p^\Lambda\,$ being the constant magnetic charges of the electric gauge fields. We work in the gauge in which the radial components $\,\mathcal{A}_r{}^\Lambda(r) \, d r\,$ are set to zero. The ansatz for the magnetic vector and the tensor field are given by
\be
\label{eq.auxiliaryansatz}
\tilde{\mathcal{A}}_0 = \tilde{\mathcal{A}}_{t \, 0}(r) \, d t - e_0  \, \frac{ \cos \sqrt{\kappa} \, \theta }{ \kappa } \, d \phi \ , \qquad
 \mathcal{B}^0 = b_0(r) \, \frac{ \sin \sqrt{\kappa} \, \theta }{ \sqrt{\kappa} } \, d \theta \wedge d \phi \ ,
\ee
where $\,e_0\,$ can be identified with a constant electric charge of $\,\mathcal{A}^0\,$ upon the use of the duality relation between electric and magnetic vectors in (\ref{EOM_auxiliary}). Furthermore, we have  made use of the tensor gauge transformations in \eqref{tensor-gauge-transf} to write only the S$^2$/H$^2$ symmetric component\footnote{\label{fn.gauge}In ref.~\cite{Klemm:2016wng}, the ansatz for the tensor field was of the form $\,\mathcal{B}_{\tiny \cite{Klemm:2016wng}}^0=\mathcal{B}_{\eqref{eq.auxiliaryansatz}}^0+d\Xi^{0}=b_0'(r)\, \tfrac{ \cos \sqrt{\kappa} \, \theta }{ \kappa } \, d r \wedge d \phi\,$ with $\,\Xi^{0} = b_0(r) \, \tfrac{ \cos \sqrt{\kappa} \, \theta }{ \kappa } \, d\phi\,$. By performing the tensor gauge transformation (\ref{tensor-gauge-transf}), the vector charges in the two gauge choices are related as $\,p^{0}(r)_{\tiny \cite{Klemm:2016wng}} = p^0_{\eqref{eq.gaugeansatz}} + \frac{1}{2} m \, b_0(r)\,$ and $\,e_{0}(r){}_{\tiny \cite{Klemm:2016wng}} = e_0{}_{\eqref{eq.auxiliaryansatz}} + \frac{1}{2} g \, b_0(r)\,$. We prefer to work with the spherically/hyperbolic symmetric form for $\,\mathcal{B}^0\,$ in (\ref{eq.auxiliaryansatz}), which is consistent with constant charges for the vector fields.}
of~$\mathcal{B}^0$.

Plugging this ansatz into the first relation of \eqref{EOM_auxiliary} implies the following constraints
\begin{equation}
\label{EOM_A0tilde}
m \, e_0 - g \, p^0 = 0 \ , \qquad
b_0'  =  e^{4 \phi + 2 \psi - 4 U} \left( g \, \mathcal{A}_t{}^0 - m \, \tilde{\mathcal{A}}_{t \, 0} \right) \ , \qquad
a'  = 0 \ , 
\end{equation}
and we can use the last one to set $\,a=0\,$. Furthermore, the U(1) current sourcing the right-hand-side of the Maxwell equation \eqref{eq.A1vector} for the $\,\mathcal{A}^1$ vector vanishes whenever $\,\zeta=\tilde \zeta=0\,$. This allows  to introduce the dual magnetic vector to $\,\tilde{\mathcal{A}}_1\,$
\be
\tilde{\mathcal{A}}_1 = \tilde{\mathcal{A}}_{t \, 1}(r) d t - e_1 \, \frac{ \cos \sqrt{\kappa} \, \theta }{ \kappa } \, d \phi\ ,
\ee
satisfying
\be\begin{split}
\label{A1_aux_vector}
d\tilde{\mathcal{A}_{1}} = \cI_{1\Lambda} * \! \mathcal{H}^\Lambda + \cR_{1\Lambda} \, \mathcal{H}^\Lambda  \ ,
\end{split}
\ee
such that the  charge $\,e_1\,$ is a constant of motion. 
Combining (\ref{A1_aux_vector}) with the second equation in (\ref{EOM_auxiliary}) we can then write duality relations between electric and magnetic vectors of the form
\be
\label{eq.vectors}
d \tilde{\mathcal{A}}_\Lambda + \frac{1}{2} \, g \, \mathcal{B}^0 \, \delta_{0 \Lambda} = \cI_{\Lambda \Sigma} * \! \mathcal{H}^\Sigma  + \cR_{\Lambda \Sigma} \, \mathcal{H}^\Sigma  \ .
\ee
Note that we do not need to solve for $\,\tilde{\mathcal{A}}_{t \, 1}\,$ as it does not enter any equation of motion. On the other hand, the integration constant $\,e_1\,$ makes and appearance in the first order equations \eqref{eq.vectors}. These read
\begin{equation}
\label{Atprim_eqs}
\begin{split}
{\mathcal{A}_t{}^0}' & =   e^{2 U - 2 \psi - 3 \varphi} \, \left[ \left( p^0 + \tfrac{1}{2}  \, m \, b_{0} \right) \, e^{6 \varphi} \, \chi^3 + 3 \, p^1 \, e^{2 \varphi} \, \chi \,  \left( 1 + e^{2 \varphi} \chi^2 \right)^2 \right. \\
&  \left. \quad  - \left( e_0 + \tfrac{1}{2} \, g \, b_{0} \right) \,  \left( 1 + e^{2 \varphi} \chi^2 \right)^3  -  e_1 \, e^{4 \varphi} \, \chi^2 \,  \left( 1 + e^{2 \varphi} \chi^2 \right)   \right]  \ , \\[2mm]
{\mathcal{A}_t{^1}}' & =   e^{2 U - 2 \psi + 3 \varphi} \, \left[ \left( p^0 + \tfrac{1}{2}  \, m \, b_{0} \right) \, \chi + 2 \, p^1 \, e^{-2 \varphi} \, \chi \, \left( 1 + 3 \, e^{2 \varphi} \, \chi^2 \right) \right. \\
&  \left.  \quad  - \left( e_0 + \tfrac{1}{2} \, g \, b_{0} \right) \, e^{-2 \varphi} \, \chi^2 \,  \left( 1 + e^{2 \varphi} \chi^2 \right)  - \frac{1}{3} \, e_1 \, e^{-2 \varphi} \,  \left( 1 + 3 \, e^{2 \varphi} \chi^2 \right)  \right] \ , \\[2mm]
{\tilde{\mathcal{A}}_{t \, 0}}{}' & =  e^{2 U - 2 \psi + 3 \varphi} \, \left[ \left( p^0 + \tfrac{1}{2}  \, m \, b_{0} \right)  + 3 \, p^1 \,  \chi^2 - \left( e_0 + \tfrac{1}{2} \, g \, b_{0} \right) \,  \chi^3 - e_1 \,  \chi \right] \ .
\end{split}
\end{equation}
The second expression in (\ref{Atprim_eqs}) allows one to integrate out $\,{\mathcal{A}}_t{^1}$ since it appears only via radial derivatives. On the other hand, the temporal components of the electric and magnetic fields $\,\mathcal{A}_{t}{}^{0}\,$ and $\,\mathcal{\tilde{A}}_{t \, 0}\,$ enter the equations of motion of the remaining fields via the combination $\,{\alpha_t^- = g \, \mathcal{A}_t{}^0 - m  \,\tilde{\mathcal{A}}_{t \, 0}}\,$.

Summarising, the spherical/hyperbolic and static ansatz we have imposed reduces the equations of motion to a system of two first-order differential equations (for $\,b_0\,$ and $\,\alpha_t^-\,$) and five second-order differential equations (for $\,\phi\,$, $\,\varphi\,$, $\,\chi\,$, $\,U\,$ and $\,\psi\,$), together with a first-order constraint coming from the radial component of the Einstein equations. The equations of motion of $\,\varphi\,$ and $\,\chi\,$ are displayed in \eqref{eq.varphieom} and \eqref{eq.chieom}. The equations of motion of $\,U\,$, $\,\psi\,$ and $\,\phi\,$ simplify to
\be
\begin{split}
\psi'' - U'' + \left( \psi' - U' \right)^2 + \phi'^2 + \frac{3}{4} \left( \varphi'^2 + e^{2 \varphi} \chi'^2 \right)  + \frac{1}{4} \, e^{4 \phi - 4 U}\, (\alpha_t^-)^2 & = 0 \ , \\
\psi'' + 2 \, \psi'^2 - e^{-2 \psi} + 2 \, e^{-2 U} \, V_{g} - \frac{1}{2} \, e^{4 \phi - 4 U}\, (\alpha_t^-)^2 & = 0 \ , \\
\phi'' + 2 \, \psi' \, \phi' - \frac{1}{2} \, e^{-2 U} \, \partial_\phi V_{g} + \frac{1}{2} \, e^{4 \phi - 4 U}\, (\alpha_t^-)^2 & = 0 \ .
\end{split}
\ee

\subsection{First-order BPS equations}

The equations of motion obtained from the Lagrangian \eqref{Lagrangian_N2} with the spherical/hyperbolic and static ansatz plugged in can be obtained from the effective one-dimensional action
\begin{equation}
\label{S_effective}
S_{1d} = \int dr \Big[ e^{2\phi} \, \big( U'^{2} - \psi'^{2} + h_{uv} \, {q^u}' {q^v}' + K_{z\bar{z}} \, z' \bar{z}'+ \frac{1}{4} \, e^{4(U-\psi)}\, \mathcal{Q}'^{\,T} \, \mathcal{H}^{-1} \, \mathcal{Q}' \big) - V_{1d}  \Big] \ ,
\end{equation}
where the primes denote derivatives with respect to the radial coordinate $\,r\,$. As pointed out in the previous section (see footnote~\ref{fn.gauge}), the ansatz for the tensor fields in \cite{Klemm:2016wng} differs from the one in (\ref{eq.auxiliaryansatz}) by a tensor gauge transformation (\ref{tensor-gauge-transf}). Consequently, our symplectic vector $\,\mathcal{Q}^{M}\,$ containing the vector charges is given by
\begin{equation}
\label{Q_model}
{\cQ^M} = \begin{pmatrix} \,\, p^0 + \frac{1}{2} \, m \, b_{0}(r) \,\, , \,\,  p^1 \,\,,\,\,  e_0 + \frac{1}{2} \, g \, b_0(r) \,\,,\,\, e_1  \,\, \end{pmatrix}^{T} \ .
\end{equation}
The matrix $\,\mathcal{H}=(\mathcal{K}^u)^{T} \, h_{uv} \, \mathcal{K}{}^{v}\,\,$ depends on the quaternionic scalars and, in our model, it takes the form
\begin{equation}
\mathcal{H}  = \frac{e^{4 \phi}}{4}  \,   
\begin{pmatrix}
\,\, m^2 & 0 & g \, m & 0 \,\, \\[1mm]
 0 & 0 & 0 & 0 \\[1mm]
 g \, m & 0 & g^2 & 0 \\[1mm]
0 & 0 & 0 & 0 
\end{pmatrix}
 \ ,
\end{equation}
where the fourth row and column are zero due to our restriction \eqref{zeta=barzeta=0}. The matrix $\,\mathcal{H}\,$ is non-invertible. This seems at odds with the appearance of $\,\mathcal{H}^{-1}\,$ in the effective action \eqref{S_effective} but, as discussed in detail in \cite{Klemm:2016wng}, the matrix $\,\mathcal{H}^{-1}\,$ is defined to satisfy the condition $\,\mathcal{H} \, \mathcal{H}^{-1} \, \mathcal{H}=\mathcal{H}\,$, which is weaker than $\,\mathcal{H}^{-1}\mathcal{H}=\mathbb{I}\,$.
Finally,  the one-dimensional potential  $\,V_{1d}\,$ is given by
\begin{equation}
\label{Vtilde}
V_{1d} = \kappa - e^{2(U-\psi)} \, V_{\textrm{BH}} - e^{-2(U-\psi)} \, V_{g} \ , 
\end{equation}
with $\,V_{\textrm{BH}}=-\frac{1}{2} \mathcal{Q}^{\,T} \, \mathcal{M} \, \mathcal{Q}\,$ being the black hole potential in $\,\mathcal{N}=2\,$ ungauged supergravity, that depends on the charges and on the scalar matrix $\,\mathcal{M}(z)\,$ in (\ref{M-matrix}).

The authors of \cite{Klemm:2016wng} also identified a real function $\,2|W|\,$ that solves the Hamilton-Jacobi equation for the effective action (\ref{S_effective}) provided a charge quantisation condition holds
\begin{equation}
\label{quant_cond}
\mathcal{Q}^{x} \, \mathcal{Q}^{x} = 1 \ ,
\end{equation}
where $\,\mathcal{Q}^{x} \equiv \left\langle  \mathcal{P}^{x} , \mathcal{Q} \right\rangle\,$. The complex function $\,W\,$ is given by
\begin{equation}
\label{W_func}
W=e^{U} (\mathcal{Z} + i \, \kappa \, e^{2(\psi-U)} \, \mathcal{L}) = |W| \, e^{i \beta} \ ,
\end{equation}
in terms of the central charge $\,\mathcal{Z}=\left\langle \mathcal{Q} , \mathcal{V} \right\rangle\,$ and a superpotential $\,\mathcal{L}=\left\langle \mathcal{Q}^{x} \mathcal{P}^{x} , \mathcal{V} \right\rangle\,$. Using $\,|W|\,$, and up to a total derivative, the effective action (\ref{S_effective}) can be written as a sum of squares yielding a set of BPS first-order equations. To integrate the BPS equations it is convenient to keep the phase $\,\beta\,$ in (\ref{W_func}) as a dynamical variable, although by its very definition is not independent of the other functions in \eqref{S_effective}. The set of BPS equations following from the effective action (\ref{S_effective}) then reads \cite{Klemm:2016wng}:
\begin{equation}
\label{BPS-eqs}
\begin{split}
U' & =  - e^{-2(\psi-U)} \,  e^{-U} \, \textrm{Re}(e^{-i\beta}\, \mathcal{Z}) - \kappa \, e^{-U} \, \textrm{Im}(e^{-i\beta}\, \mathcal{L}) \ , \\[2mm]
\psi' & =  - 2 \, \kappa \, e^{-U} \, \textrm{Im}(e^{-i\beta}\, \mathcal{L}) \  , \\[2mm]
\mathcal{V}' & =  e^{i\beta} \, e^{-2(\psi-U)} \,  e^{-U} \, (-\frac{1}{2} \, \Omega \, \mathcal{M} \, \mathcal{Q} - \frac{i}{2} \, \mathcal{Q} + \mathcal{Z} \, \bar{\mathcal{V}}  ) \\
& \quad - \, i \, \kappa \, e^{i\beta} \, e^{-U} (-\frac{1}{2} \, \Omega  \, \mathcal{M} \, \mathcal{P}^{x} \mathcal{Q}^{x} - \frac{i}{2} \, \mathcal{P}^{x}\, \mathcal{Q}^{x} + \mathcal{L} \, \bar{\mathcal{V}}  ) - \, i \, A_{r} \, \mathcal{V} \ , \\[2mm]
{q^u}' & =  \kappa \, e^{-U} \, h^{uv} \,  \textrm{Im}(e^{-i\beta}\, \partial_{v} \mathcal{L})  \ , \\[2mm]
\mathcal{Q}' & =  - 4 \, e^{2 (\psi - U)} e^{-U}  \mathcal{H} \, \Omega \, \textrm{Re}(e^{-i\beta}\, \mathcal{V}) \ , \\[2mm]
\mathcal{\beta}' & =   2 \, \kappa \, e^{-U} \, \textrm{Re}(e^{-i\beta}\, \mathcal{L}) - A_{r}  \ ,
\end{split}
\end{equation}
where $\,A_{r} = \textrm{Im}(z'\partial_{z}K)= - \frac{3}{2} \, e^{\varphi (r)} \, \chi '(r)\,$ is the U(1) K\"ahler connection in $\,\mathcal{M}_{\textrm{SK}}\,$. The  system \eqref{BPS-eqs} must be supplemented with the charge quantisation condition in (\ref{quant_cond}), the expression of the phase $\,\beta\,$ as a function of the other scalars dictated by \eqref{W_func}, and with a set of additional constraints
\begin{equation}
\label{extra_constraints}
\mathcal{H} \, \Omega \, \mathcal{Q} = 0 \ , \qquad
h_{uv} \, \mathcal{K}_{M}{}^{u} \, {q^{v}}' = 0 \ , \qquad
\mathcal{H} \, \Omega \, \mathcal{A}_{t} = 2 \, e^{U} \, \mathcal{H} \, \Omega \, \textrm{Re}(e^{-i\beta} \mathcal{V}) \ ,
\end{equation}
arising as compatibility conditions with the original (unreduced) equations of motion of the vector fields. In a nutshell, the first expression in \eqref{extra_constraints} corresponds to the first condition in \eqref{EOM_A0tilde}, the second expression in  \eqref{extra_constraints} is imposed by the vector equations of motion subjected to spherical/hyperbolic symmetry and corresponds to the last condition in \eqref{EOM_A0tilde}. The third equation allows one to express $\,\alpha^-_t\,$ in terms of the scalars of the theory, therefore eliminating all explicit appearances of the vectors in the original Lagrangian from the BPS equations.

As a closing remark, the set of BPS equations \eqref{BPS-eqs} is invariant under a constant shift of the radial coordinate, as well as under a rescaling of the radial coordinate and metric functions of the form
\be
\label{rescaling_sym}
 r \to \lambda\, r \ , \qquad e^{U} \to  \lambda \, e^U \ , \qquad e^{\psi-U} \to e^{\psi-U} \ .
 \ee

\section{Black holes and BPS flows}

In this section we present the attractor equations for the near-horizon region of BPS black holes in the $\,\mathcal{N}=2\,$ supergravity model we are investigating. Then we find BPS black hole solutions for which the scalar fields both in the vector multiplet and the universal hypermultiplet vary along the radial coordinate. The generic solutions interpolate between a unique $\,\textrm{AdS}_{2} \times \textrm{H}^{2}\,$ geometry in the near-horizon region and the domain-wall $\,\textrm{DW}_{4}\,$ (four-dimensional) description of the D2-brane at $\, r \rightarrow \infty\,$. However, special behaviours at $\, r \rightarrow \infty\,$ also occur when the boundary conditions at the horizon are fine tuned. All the plots presented in this section have been generated with $\, g = m = 1 \,$, which can always be achieved by a rescaling of the fields.

\subsection{Near-horizon region and attractor equations}

The near-horizon geometry of an extremal four-dimensional black hole is given by $\,\textrm{AdS}_{2} \times \Sigma_{2}\,$, with $\,{\Sigma_{2}=\{\textrm{S}^2, \, \textrm{H}^2\}}\,$. The functions $\,e^{U(r)}\,$ and $\,e^{\psi(r)}\,$ in the metric (\ref{metric_general})  take the form
\begin{equation}
\label{U-psi-horizon}
e^{2U} = \frac{r^2}{L^2_{\textrm{AdS}_{2}}}\ , \qquad
e^{2(\psi-U)} = L^{2}_{\Sigma_{2}} \ ,
\end{equation}
where $\,L_{\textrm{AdS}_{2}}\,$ and $\,L_{\Sigma_{2}}\,$ are the curvature radii of the AdS$_{2}$ and $\,\Sigma_{2}\,$ factors of the $\,\textrm{AdS}_{2} \times \Sigma_{2}\,$ near-horizon geometry. In the parameterisation  \eqref{U-psi-horizon} we have shifted the radial coordinate $\,r\,$ to place the horizon  at $\,r_{h}=0\,$. Using the equations for $\,U'\,$ and $\,\psi'\,$ in (\ref{BPS-eqs}), and plugging in the functions (\ref{U-psi-horizon}), one obtains $\, e^{-U} (\mathcal{Z} + i \, \kappa \, L_{\Sigma_{2}}^2 \, \mathcal{L}) = 0 \,$. Since this equality has to hold for any value of the radius in the $\,\textrm{AdS}_{2} \times \Sigma_{2}\,$ fixed point, it follows that
\be
\label{Z=L_eq}
\mathcal{Z} + i \, \kappa \, L_{\Sigma_{2}}^2 \, \mathcal{L} =0 \ .
\ee
Assuming that the scalars enter the horizon as constants, \textit{i.e.} $\,z'={q^{u}}'=0\,$, it follows from (\ref{BPS-eqs}) that $\,\beta'=0\,$ and $\,\mathcal{Q}'=0\,$. Moreover, it can be shown from (\ref{Z=L_eq}) and the first relation in \eqref{extra_constraints} that $\,\left\langle  \mathcal{K}^{u} , \mathcal{V} \right\rangle=0\,$. All these consequences of the $\,\textrm{AdS}_{2} \times \Sigma_{2}\,$ form of the metric imply that the BPS equations (\ref{BPS-eqs}) can be rewritten as the set attractor equations derived in \cite{Klemm:2016wng}
\begin{equation}
\label{attractor_eqs}
\begin{split}
\mathcal{Q} & =   \kappa \, L_{\Sigma_{2}}^{2} \, \Omega \, \mathcal{M} \, \mathcal{Q} ^{x} \, \mathcal{P}^{x} - 4 \, \textrm{Im}(\bar{\mathcal{Z}} \, \mathcal{V}) \ , \\
\dfrac{L_{\Sigma_{2}}^{2}}{L_{\textrm{AdS}_{2}}} & =  -2 \, \mathcal{Z} \, e^{-i \beta} \ , \\[2mm]
\left\langle  \mathcal{K}^{u} , \mathcal{V} \right\rangle & =  0 \ ,
\end{split}
\end{equation}
where it is understood that all scalars and $\,b_0\,$ are evaluated at the horizon.
As for the general BPS equations, the charge quantisation condition (\ref{quant_cond}) and the additional constraints (\ref{extra_constraints}) must be imposed. The latter constraint also imposes $\,\mathcal{H} \, \Omega \, \mathcal{A}_{t}=0\,$, implying that in the $\,\textrm{AdS}_{2} \times \Sigma_{2}\,$ region $\, g\, {\mathcal{A}_t}^0 = m\, \tilde{\mathcal{A}}_{t \, 0} \,$ and $\,{\mathcal{A}_t}^1=0\,$.

Let us characterise the near-horizon geometries in the model arising from the reduction of the massive IIA theory on the six-sphere. First of all, since $\,\mathcal{Q}'(r_h)=0\,$, it follows from \eqref{Q_model} that
\be
{b_0}'(r_h)=0 \ .
\ee
The (quadratic) charge quantisation condition (\ref{quant_cond}) reduces in this case to
\begin{equation}
\label{QQ_model}
p^1 \, \left[ \, 1 \, + \, \frac{e^{2 \phi }}{4} \, \left(\zeta ^2+\tilde{\zeta}^2 \right)  \, \right]  = \pm \frac{1}{3g} \ ,
\end{equation}
where we  have  made use of the first constraint in (\ref{EOM_A0tilde}). Here we are reinstating temporarily the scalars $\,\zeta\,$ and $\,\tilde \zeta\,$ to show explicitly how the attractor equations set them to zero. This is seen from the last expression in (\ref{attractor_eqs}), which in particular does not involve the charges $\,\mathcal{Q}\,$. In our specific model this equation  imposes  
\begin{equation}
\label{N=2_point}
e^{\varphi_{h}} = \frac{2}{\sqrt{3}} \left( \frac{g}{m} \right)^{\frac{1}{3}}
 \ , \qquad
\chi_{h}=-\frac{1}{2} \,  \left( \frac{g}{m} \right)^{-\frac{1}{3}}
 \ , \qquad
\zeta_{h} = \tilde{\zeta}_{h} = 0 \ ,
\end{equation}
and fixes all the values of the scalars at the horizon but $\,\phi_{h}\,$ in terms of the gauging parameters. 
Substituting  (\ref{N=2_point}) into the charge quantisation condition (\ref{QQ_model}) gives
\begin{equation}
p^{1}=\pm \frac{1}{3g} \ .
\end{equation}
Plugging these results into the first and second equations in (\ref{attractor_eqs}) produces a set of algebraic relations. The system has a solution only if  $\,\kappa=-1\,$ (hyperbolic horizon) and the scalars, charges and radii take the values
\begin{equation}
\label{horizon_fields}
\begin{split}
e^{\varphi_{h}} = \frac{2}{\sqrt{3}} \, \left( \frac{g}{m} \right)^{\frac{1}{3}}
\ , \quad
\chi_{h}=-\frac{1}{2} \,  \left( \frac{g}{m} \right)^{-\frac{1}{3}}
& \ , \quad
e^{\phi_{h}} = \sqrt{2} \, \left( \frac{g}{m} \right)^{\frac{1}{3}}
\ , \quad
a_{h}=\zeta_{h} = \tilde{\zeta}_{h} = 0 \ ,  \\
p^{0}+\frac{1}{2} \, m \, b^{h}_{0}=\pm \, \frac{1}{6}  \, m^{\frac{2}{3}} \, g^{-\frac{5}{3}}
& \ , \quad 
e_{0}+\frac{1}{2} \, g \, b^{h}_{0}=\pm \, \frac{1}{6} \, m^{-\frac{1}{3}} \, g^{-\frac{2}{3}} \ , \\
p^{1}=\mp \, \frac{1}{3} \, g^{-1}
& \ , \quad 
e_{1} = \pm \, \frac{1}{2} \, m^{\frac{1}{3}} \, g^{-\frac{4}{3}} \ ,  \\
L^2_{\textrm{AdS}_{2}}= \frac{1}{4\, \sqrt{3}} \,  m^{\frac{1}{3}} \, g^{-\frac{7}{3}}
& \ , \quad
L^2_{\textrm{H}^{2}}= \frac{1}{2\, \sqrt{3}} \,  m^{\frac{1}{3}} \, g^{-\frac{7}{3}} \ .
\end{split}
\end{equation}
The two horizon configurations are related to each other by an overall change in the sign of the charges $\,\mathcal{Q}_{h} \rightarrow -\mathcal{Q}_{h}\,$. Moreover, using the definition of the phase $\,\beta\,$ given in \eqref{W_func}, one finds that $\,\beta^h = \frac{\pi}{3} \mp \frac{\pi}{2}\,$. From now on we select the first of these configurations, namely, the one with $\,\beta^{h}=-\pi/6\,$.

\subsection{Asymptotically AdS$_{4}$ solutions with charges}

The same configuration of the scalar fields that we have found in the analysis of the attractor equations can be seen to extremise the scalar potential $\,V_{g}\,$ in (\ref{Vg_model}). In absence of charges, this configuration supports an $\,{\textrm{AdS}_{4} \times  \textrm{S}^{6}}\,$ solution of massive IIA supergravity preserving $\,\mathcal{N}=2\,$ supersymmetry and $\,\textrm{SU}(3) \times \textrm{U}(1)\,$ symmetry \cite{Guarino:2015jca,Varela:2015uca}. As a consequence of the spherical/hyperbolic symmetry, the metric functions depend explicitly on $\,\kappa\,$ and take the form
\be
e^{2U}= \kappa + \frac{ \tilde{r}^2 }{L^{2}_{\textrm{AdS}_{4}} } \ , \qquad e^{2(\psi-U)} = \tilde{r}^2 \ , 
\ee
with $\,L^{2}_{\textrm{AdS}_{4}} =\frac{3}{|V^{*}_{g}|}=\frac{1}{\sqrt{3}} \, m^{\frac{1}{3}} \, g^{-\frac{7}{3}}\,$ and $\,V^{*}_{g}\,$ being the value of the potential (\ref{Vg_model}) at the extremum. Here we are denoting the radial coordinate as $\,\tilde r\,$ since, as we show below, it is shifted by a constant with respect to the one used in the previous section.

Since the set of BPS equations \eqref{BPS-eqs} requires the quantisation condition \eqref{quant_cond} to be satisfied, it is clear that this solution is not captured in the present setup. However it can be shown that, in the presence of charges, there is a Reissner--Nordstr\"om-AdS like solution with the same value for the scalars \cite{Caldarelli:1998hg} and with
\begin{equation}
\label{charged_solus}
e^{2U} =  \kappa  + \frac{f(\mathcal{Q})}{\tilde{r}^2} +  \frac{ \tilde{r}^2 }{L^{2}_{\textrm{AdS}_{4}} } \ , 
\qquad
e^{2(\psi-U)} =\tilde{r}^2 \ .
\end{equation}
Substituting (\ref{charged_solus}) into the BPS equations (\ref{BPS-eqs}), one finds a one-parameter family of solutions with charges
\begin{equation}
\label{charged_solus_2}
\begin{split}
p^{0} + \tfrac{1}{2} \, m \, b_{0} & = -\frac{1}{3} \, m^{\frac{1}{3}} \, g^{-\frac{1}{3}} \, e_{1}  - \frac{\kappa}{3} \, m^{\frac{2}{3}} \, g^{-\frac{5}{3}}\ , \qquad p^{1} =  \frac{\kappa}{3\,g}  \ ,  \\
e_{0} + \tfrac{1}{2} \, g \, b_{0} & = -\frac{1}{3} \, m^{-\frac{2}{3}} \, g^{\frac{2}{3}} \, e_{1}  - \frac{\kappa}{3} \, m^{-\frac{1}{3}} \, g^{-\frac{2}{3}}
\ , \quad\,\,\,
e_{1} =  \textrm{free} \ ,
\end{split}
\end{equation}
which yields a function $\,f(\mathcal{Q})\,$ in (\ref{charged_solus}) of the form
\begin{equation}
f(\mathcal{Q})= \frac{1}{3\sqrt{3}} \left( \kappa\, m^{\frac{1}{6}} \, g^{-\frac{7}{6}} - m^{-\frac{1}{6}} \, g^{\frac{1}{6}} \frac{}{} \, e_{1}  \right)^2 + \frac{\kappa}{\sqrt{3}\, g} \, e_{1} \ .
\end{equation}
This one-parameter family  of solutions corresponds to an asymptotically AdS$_4$ geometry with non-trivial charges turned on. Near the origin, $\tilde{r}=0$, the solution gives rise to a naked singularity.  The family admits a non-extremal generalisation by adding to the metric function $\,e^{2U}\,$ in \eqref{charged_solus} a mass term of the form $-2M/\tilde{r}$. With this the metric is a solution of the second-order equations of motion in appendix \ref{App:EOMs} (but not of the BPS equations), and the geometry in the IR is regularised by a horizon. This indicates that the naked singularities of \eqref{charged_solus} are of the good type in the classification of \cite{Gubser:2000nd}.
There is a particular case of the BPS solution \eqref{charged_solus_2} with
\be
\label{charges_sol_analytic}
\kappa=-1 \ , \qquad
e_1 = \frac{1}{2} \, m^{\frac{1}{3}} \, g^{-\frac{4}{3}} \ ,
\ee
which connects with the attractor solution in (\ref{horizon_fields}). It corresponds to an extremal Reissner--Nordstr\"om black hole solution with $\,\textrm{AdS}_2 \times \textrm{H}^{2}\,$ geometry in the IR. This choice of $\,e_1\,$ charge yields a function $\,f(\mathcal{Q})\,$ in (\ref{charged_solus}) of the form
\begin{equation}
\label{metric_sol_analytic}
f(\mathcal{Q}) = \frac{m^{\frac{1}{3}} \, g^{-\frac{7}{3}}}{4 \, \sqrt{3}} \qquad  \Rightarrow \qquad e^{2U} = \left( \frac{\tilde{r}}{ L_{\textrm{AdS}_{4}} } - \frac{ L_{\textrm{AdS}_{4}} }{2\, \tilde{r}} \right)^2 \ ,
\end{equation}
with the horizon located at $\,\tilde{r}^2_h=\tfrac{1}{2\sqrt{3}} \, m^{\frac{1}{3}} \, g^{-\frac{7}{3}}$.

\subsection{BPS flows from the $\,\mathrm{DW}_4\,$ to $\textrm{AdS}_{2} \times \textrm{H}^{2}$}

\begin{figure}[t!]
\begin{center}
\includegraphics[width=0.7\textwidth]{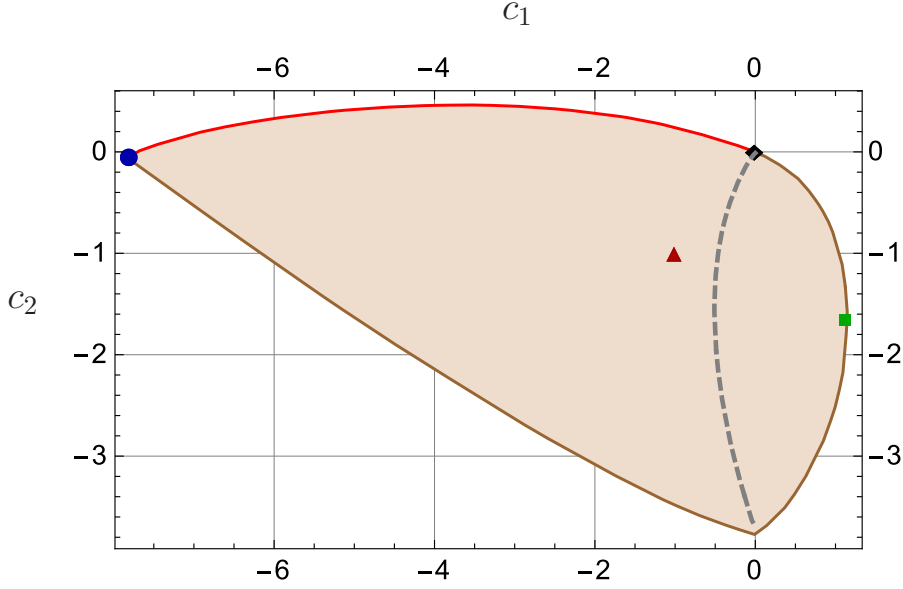}
\put(-150,214){\Large $c_1$ }
\put(-335,105){\Large $c_2$}
\caption{Plot of the two-dimensional parameter space $\,(c_{1},c_{2})\,$ of BPS solutions (shaded area) interpolating between the $\,{\textrm{AdS}_{2} \times \textrm{H}^{2}}\,$ geometry in the IR and the $\,\mathrm{DW}_4\,$ solution in the UV.
\label{fig.region}}
\end{center}
\end{figure}

We have shown that the attractor equations \eqref{attractor_eqs} select a unique configuration of charges (modulo a $\mathbb{Z}_{2}$ transformation) and scalar fields, given in (\ref{horizon_fields}), such that a horizon with hyperbolic symmetry exists. This $\,{\textrm{AdS}_{2} \times \textrm{H}^{2}}\,$ geometry in the IR can be reached from a charged $\,\textrm{AdS}_{4}\,$ geometry in the UV yielding the extremal BH solution in (\ref{charges_sol_analytic})-(\ref{metric_sol_analytic}) with constant scalars. In this section we construct numerically more BPS solutions, and show that the analytic BH-AdS geometry corresponds to a very special point within a two-dimensional parameter space of  configurations. These solutions generically interpolate between an $\,{\textrm{AdS}_{2} \times \textrm{H}^{2}}\,$ geometry in the IR and a $\,\mathrm{DW}_4\,$ domain-wall geometry governed by the D2-brane in the UV~(see Figure~\ref{fig.region}). 

To understand how the UV geometry  is dictated by the D2-brane,  let us recall the form of such a solution in massless IIA supergravity.
This is given by a metric (in Einstein frame) and a dilaton $\,e^{\hat{\Phi}}\,$ of the form
\begin{equation}
\label{D2_brane_massless}
d\hat{s}_{10}^2 = e^{\frac{3}{4}\phi} \left( -e^{2U} dt^2 + e^{-2U} dr^2 + e^{2(\psi-U)} ds_{\Sigma_{2}}^2 \right) + g^{-2} e^{-\frac{1}{4}\phi} ds^2_{\textrm{S}^{6}} \ , \qquad e^{\hat{\Phi}} = e^{\frac{5}{2}\phi} \ .
\end{equation}
In addition, there is a four-form flux $\,\hat{F}_{(4)} = 5 \, g \, e^{\phi} \, e^{2(\psi-U)} \, dt \wedge dr \wedge d\Sigma_{2} \,$ that is electrically sourced by the D2-brane. The leading UV dependence on the radial coordinate of the different functions is given by
\begin{equation}
\label{D2-scaling}
e^{2 U} \sim  r^{\frac{7}{4}}
\ , \qquad
e^{2 (\psi-U)}  \sim r^{\frac{7}{4}}
 \ , \qquad 
e^{\phi} \sim  r^{-\frac{1}{4}} \ .
\end{equation}
The four-dimensional $\,\mathrm{DW}_4\,$ domain-wall description of the D2-brane in (\ref{D2-scaling}) is an exact solution to the equations of motion in appendix~\ref{App:EOMs} only if one sets the charges and the Romans' mass to zero, takes $\,\Sigma_{2}=\mathbb{R}^2\,$, and restricts the scalars to the SO(7)-invariant sector: $\,\chi=0\,$ and $\,e^{\varphi}=e^{\phi}\,$. When turning on the Romans' mass and/or the charges and/or a non-trivial $\,\Sigma_{2}\,$, the metric and dilaton fields in (\ref{D2-scaling}) are no longer an exact solution of the theory. Their presence necessarily adds corrections to the behaviour in (\ref{D2-scaling}) which are suppressed as one approaches the boundary at $\,{r \to \infty}\,$ (see appendix~\ref{App:UV_expansion} for an explicit expansion). Taking as an example the case of the Romans' mass, this can be understood from the potential of the corresponding four-dimensional gauged supergravity or from the fermion mass terms entering the supersymmetry transformations obtained upon reduction on $\,\textrm{S}^{6}\,$. In both cases the Romans' mass parameter appears dressed up with a function of the scalars that suppresses its contribution near the boundary. A similar effect occurs in the case of non-trivial charges: they are dressed up with functions of the scalars that make their induced corrections subleading near the boundary. 

Furthermore, perturbing the BPS equations around the $\,\mathrm{DW}_4\,$ geometry shows that only relevant deformations are turned on \cite{Guarino:2016ynd}. For this reason, the D2-brane solution of the massless IIA theory generically governs the UV asymptotics also in the massive setup with finite charges. In addition, having a solution whose UV is governed by the $\,\mathrm{DW}_4\,$  configuration necessarily implies a running of the dilaton $\,e^\phi\,$ belonging to the universal hypermultiplet. This implies that all the solutions that we describe in this section contain running hyperscalars.

\begin{figure}[t!]
\begin{center}
\hspace{-4mm}
\begin{subfigure}[b]{0.31\textwidth}
\includegraphics[width=\textwidth]{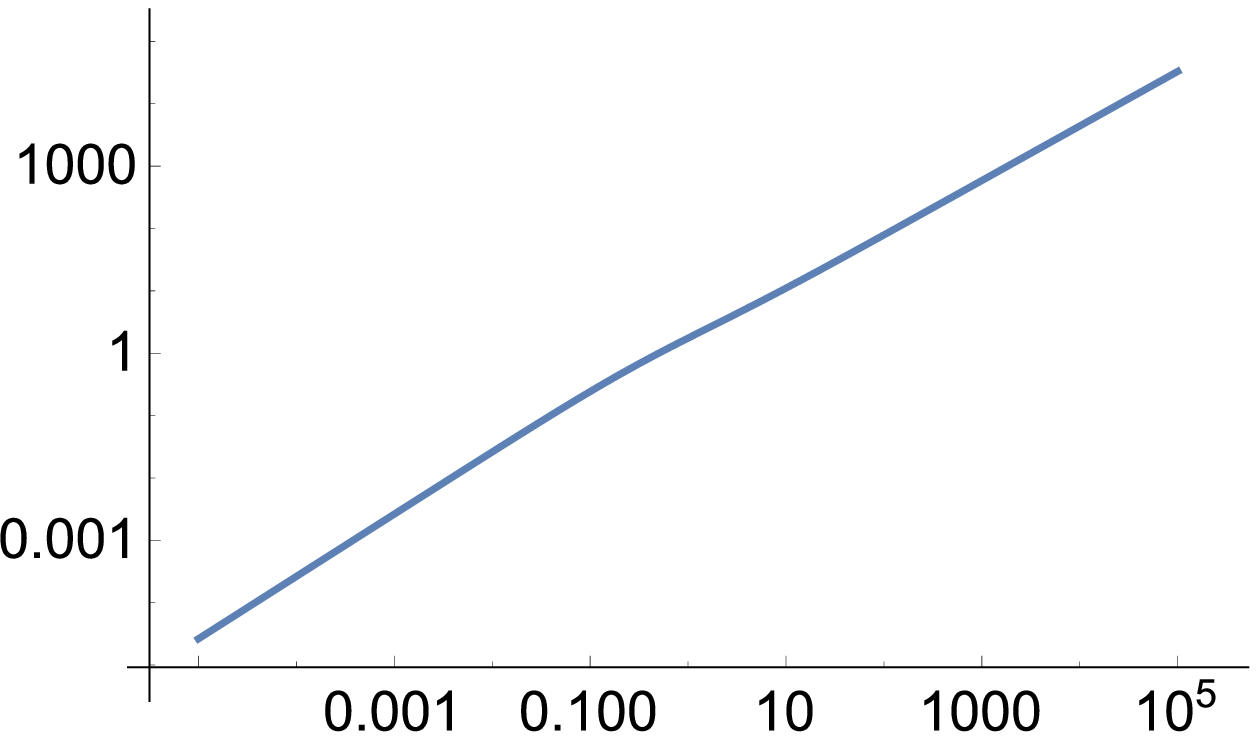}
\put(3,5){\small $r$}
\put(-130,90){\small $e^U$}
\end{subfigure}
~~
\begin{subfigure}[b]{0.31\textwidth}
\includegraphics[width=\textwidth]{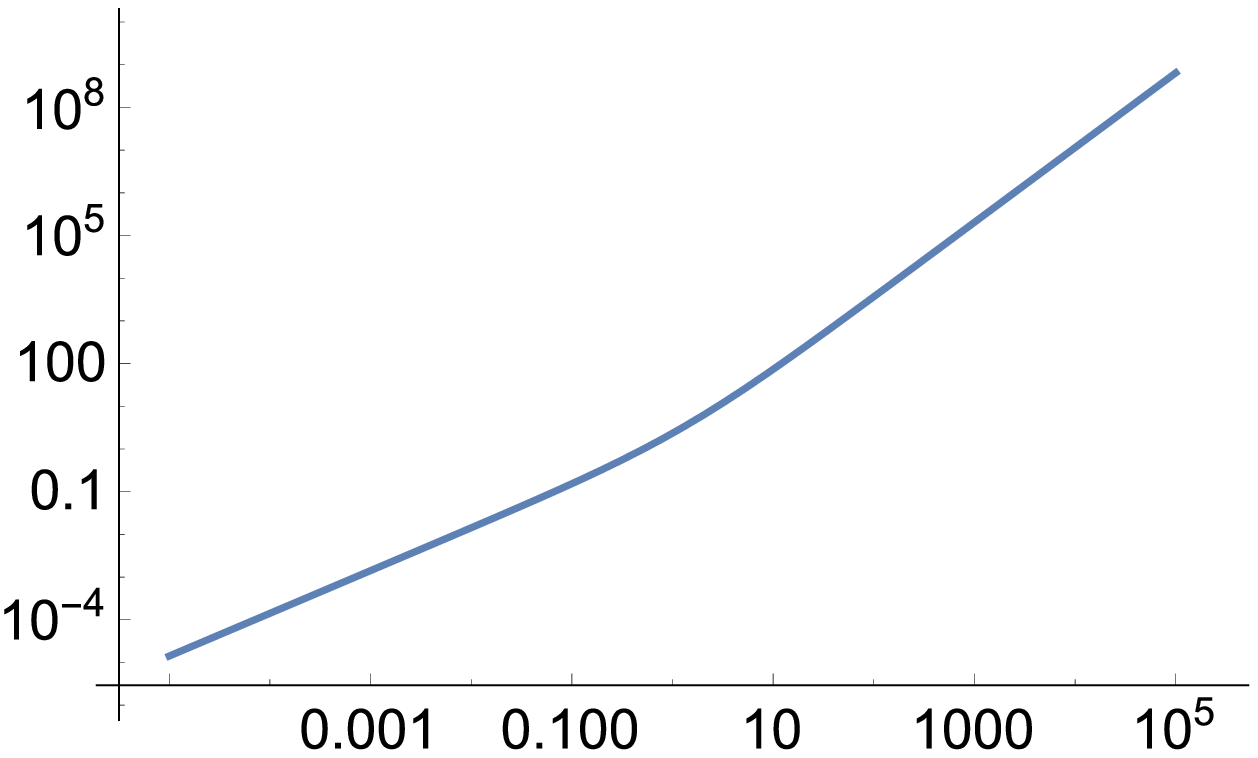}
\put(3,5){\small $r$}
\put(-130,90){\small $e^\psi$}
\end{subfigure}
~~
\begin{subfigure}[b]{0.31\textwidth}
\includegraphics[width=\textwidth]{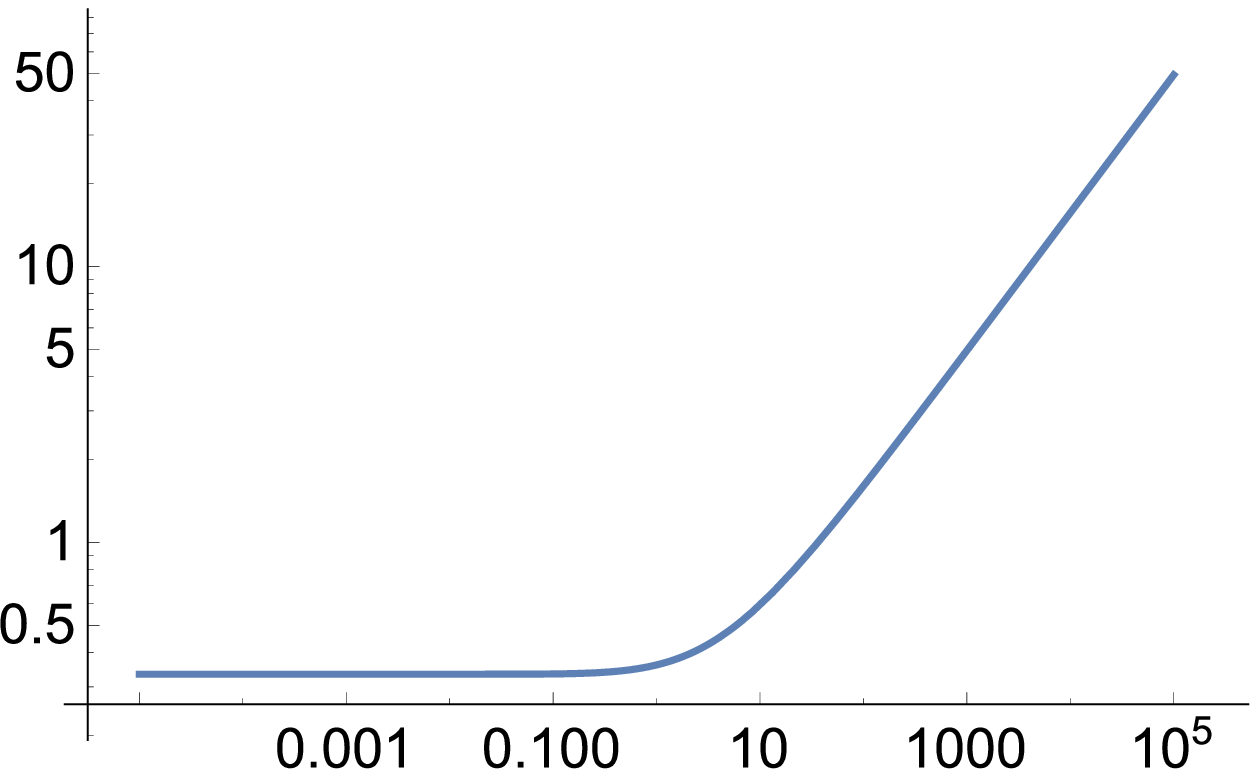}
\put(3,5){\small $r$}
\put(-135,90){\small $b_0$}
\end{subfigure}
\\[2mm]
\hspace{-4mm}
\begin{subfigure}[b]{0.31\textwidth}
\includegraphics[width=\textwidth]{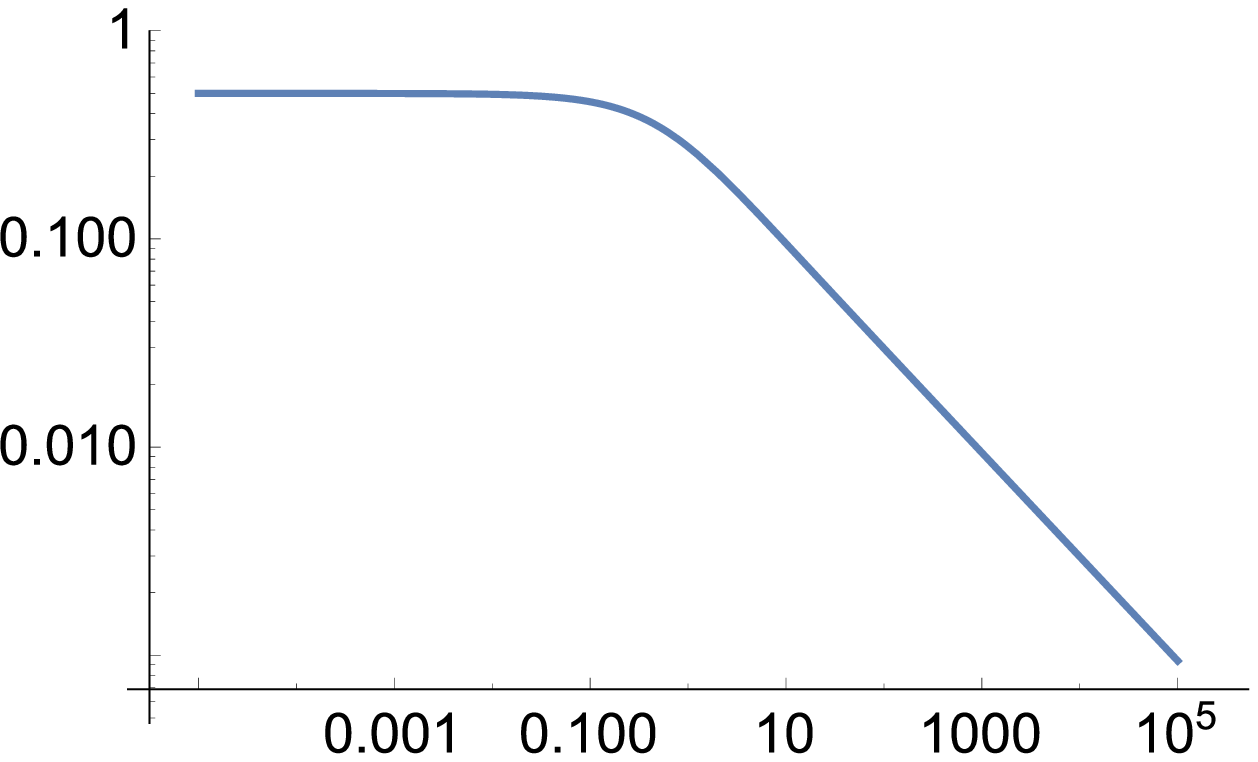}
\put(3,5){\small $r$}
\put(-135,90){\small $- \chi$}
\end{subfigure}
~~
\begin{subfigure}[b]{0.31\textwidth}
\includegraphics[width=\textwidth]{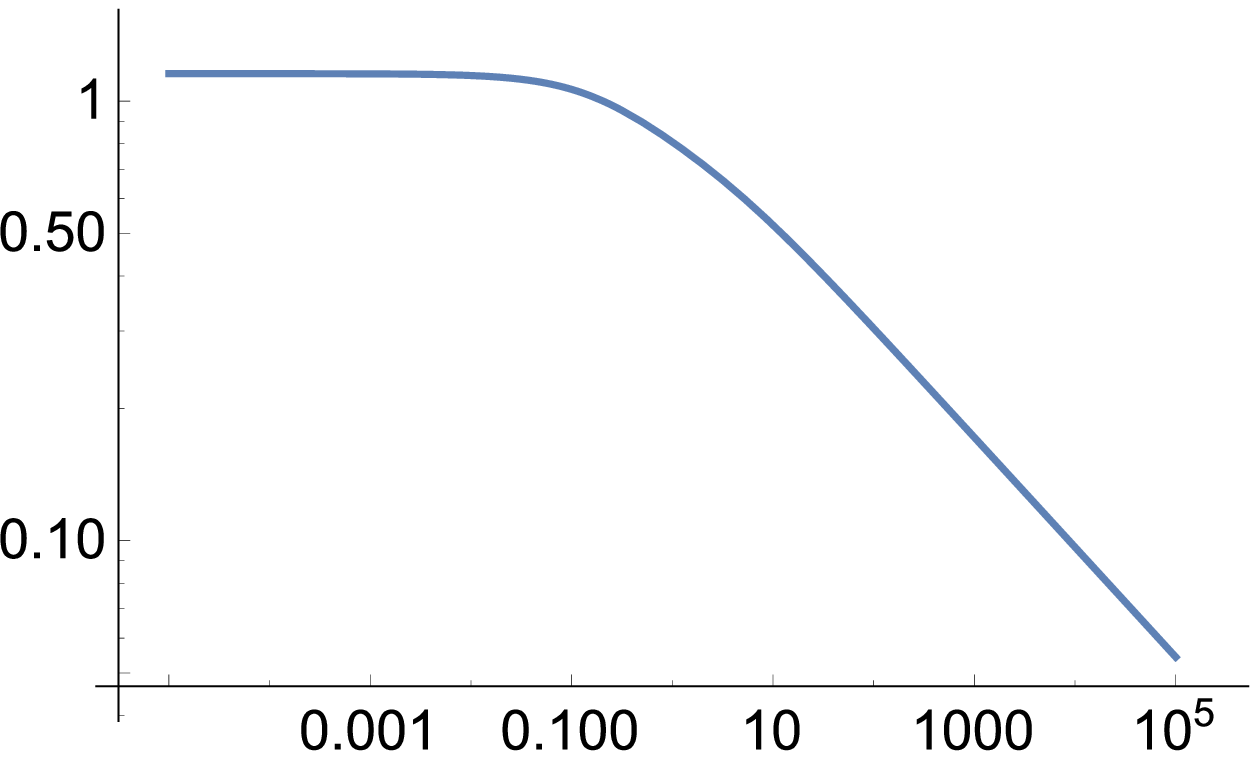}
\put(3,5){\small $r$}
\put(-135,90){\small $e^\varphi$}
\end{subfigure}
~~
\begin{subfigure}[b]{0.31\textwidth}
\includegraphics[width=\textwidth]{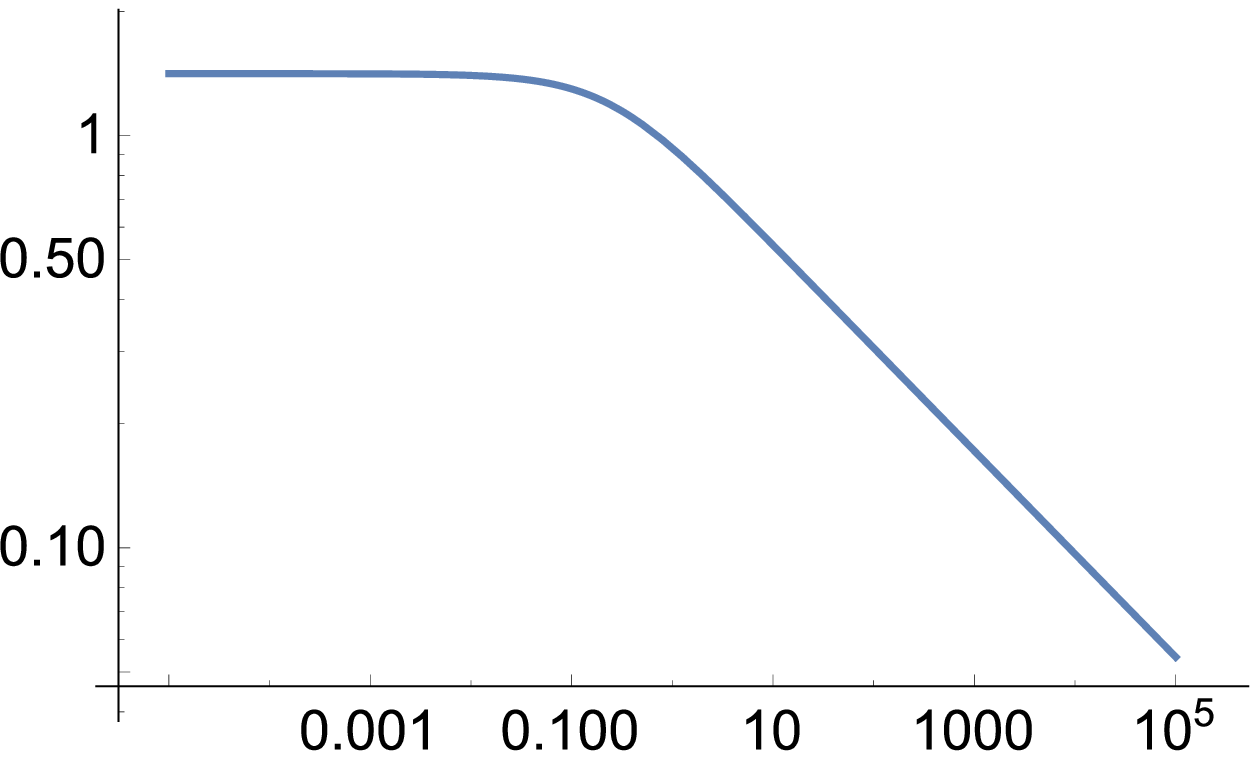}
\put(3,5){\small $r$}
\put(-135,90){\small $e^\phi$}
\end{subfigure}
\caption{Plots of the metric functions, scalars and tensor field profiles as a function of the radial coordinate. The numerical integration was performed with $\,(c_1,c_2)=(-1,-1)\,$. 
\label{fig.generic}}
\end{center}
\end{figure}

In order to solve the BPS equations, we shoot numerically from the extremal horizon. To impose appropriate boundary conditions, we  first identify the irrelevant perturbations around the unique $\,\textrm{AdS}_{2} \times \textrm{H}^{2}\,$ solution given by the metric and fields in \eqref{U-psi-horizon} and \eqref{horizon_fields}. Expanding the BPS equations \eqref{BPS-eqs} near the horizon at $\,r=0\,$, one finds the following regular corrections to the metric and field functions: 
\be
\label{eq.irrelevantIR}
\begin{split}
e^U & \simeq \frac{r}{L_{\textrm{AdS}_{2}}} \left( 1 - \lambda \, r \right) \ , \qquad e^{\psi-U} \simeq L_{\textrm{H}^{2}} \left( 1 +2 \, \lambda \, r \right) \ , \\
\chi &  \simeq \chi_h \left( 1 + c_1 \, r \right) \ , \qquad e^\varphi  \simeq e^{\varphi_h} \left( 1 + c_2 \, r \right) \ , \qquad e^\phi  \simeq e^{\phi_h} \, \Big( \, 1 + \tfrac{1}{4} \, (c_1+3 \, c_2) \, r \, \Big) \ , \\
 b_0 & \simeq b_0^h - \tfrac{1}{2} \, (c_1 - c_2) \, m^{-\frac{1}{3}} \, g^{-\frac{5}{3}} \, r  \ , \qquad \beta \simeq \beta^h  - \frac{\sqrt{3}}{2} \,  c_1 \, r \ .
\end{split}
\ee
Therefore, there are three parameters $\,\lambda\,$ and $\,(c_{1},c_{2})\,$ that describe the irrelevant deformations around the $\,\textrm{AdS}_2\times \textrm{H}^{2}\,$ solution. The first one, $\,\lambda\,$, describes the perturbation of the metric functions and can be set to any (positive) value by virtue of the scaling symmetry \eqref{rescaling_sym} of the BPS equations. We choose\footnote{There is also the possibility to set $\,\lambda=0\,$. In this case we have not found any regular solution to the equations of motion besides the trivial $\,\lambda=c_1=c_2=0\,$ solution that does not flow away from the IR fixed point.}
\be
\lambda = \tfrac{1}{\sqrt{2}} \, 3^{\frac{1}{4}} \ , 
\ee
as in the asymptotically AdS$_{4}$ solution (\ref{charges_sol_analytic})-(\ref{metric_sol_analytic}). The remaining parameters $\,(c_1,c_2)\,$ parameterise the irrelevant deformations describing how the solutions arrive at the $\,{\textrm{AdS}_{2} \times \textrm{H}^{2}}\,$ geometry in the IR.

We have performed a numerical scan of $\,10^6\,$ points in the $\,(c_1,c_2)$-plane within the range $\,-100 \le c_{1,2} \le 100\,$. The result is depicted in Figure~\ref{fig.region}, which we explain now in some detail. The shaded region corresponds to regular BPS configurations that interpolate between the $\,{\textrm{AdS}_{2} \times \textrm{H}^{2}}\,$ solution \eqref{U-psi-horizon} and \eqref{horizon_fields} in the IR, and flow to the DW$_{4}$ solution (\ref{D2-scaling}) in the UV. All these configurations have the same behaviour at large $\,r\,$ given by (\ref{D2-scaling}) together with
\be
\label{D2-scaling_massive}
\begin{split}
\chi & \sim -r^{-1/2} \ , \qquad e^\varphi \simeq e^\phi  
\ , \qquad
b_0 \sim r^{1/2} \ , \qquad \beta \sim - r^{-3/4} \ .
\end{split}
\ee
In (\ref{D2-scaling_massive}) we are omitting corrections that fall off at $r\to\infty$ with coefficients depending on $\,(c_1,c_2)\,$ that can be found in appendix~\ref{App:UV_expansion}. Importantly for the D2-brane interpretation, the two dilatons $e^\varphi$ and $e^\phi$ become identified asymptotically and the axion $\,\chi\,$ goes to zero faster than the dilatons as $\,r\,$ increases. The  BPS solution with $\,(c_{1},c_{2})=(-1,-1)\,$ is represented in Figure~\ref{fig.region} by a (red) triangle, and the profiles for the corresponding fields are shown in Figure~\ref{fig.generic}. Note that, despite this solution having $\,c_1=c_2\,$, the function $\,b_0\,$ still flows non-trivially as it receives a correction at a larger order than the one given in \eqref{eq.irrelevantIR}. 

The divergent behaviour of the non-propagating tensor field $\,b_{0}\,$ in (\ref{D2-scaling_massive}) renders some of the charges in (\ref{Q_model}) divergent but does not spoil the finiteness of the on-shell action, thus indicating that this mode does not carry infinite energy at the boundary.\footnote{Plugging (\ref{D2-scaling}) and (\ref{D2-scaling_massive}) into (\ref{Atprim_eqs}) it is straightforward to show that the norm of the gauge potentials $\,|\mathcal{A}^{0,1}|^2\,$ goes to zero near the boundary. The same holds for the auxiliary fields $\,|\tilde{\mathcal{A}}_{0}|^2\,$ and $\,|\mathcal{B}^{0}|^2\,$.} This is also supported by the metric asymptoting to the DW$_4$ solution in (\ref{D2-scaling}). Nonetheless, it is possible to tune the values of $\,(c_1,c_2)\,$ to find solutions such that $\,b_0\,$ approaches a constant value when $\,r\to\infty\,$ (see appendix~\ref{App:UV_expansion}). We have denoted the locus of such parameters with the (grey) dashed line in Figure~\ref{fig.region}.

\begin{figure}[t!]
\begin{center}
\begin{subfigure}[b]{0.4\textwidth}
\includegraphics[width=\textwidth]{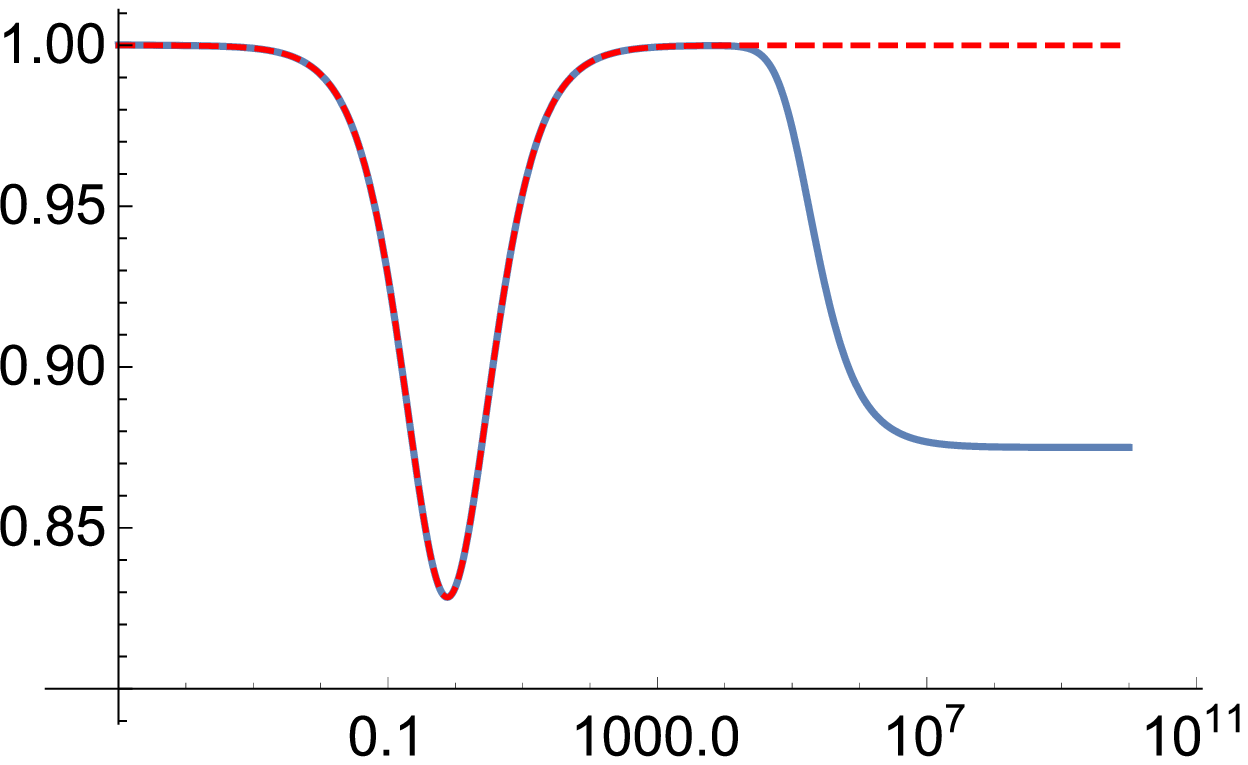}
\put(3,5){ $r$}
\put(-175,120){ $r \, U'$}
\end{subfigure}\vspace{2mm}
~~~~~~
\begin{subfigure}[b]{0.4\textwidth}
\includegraphics[width=\textwidth]{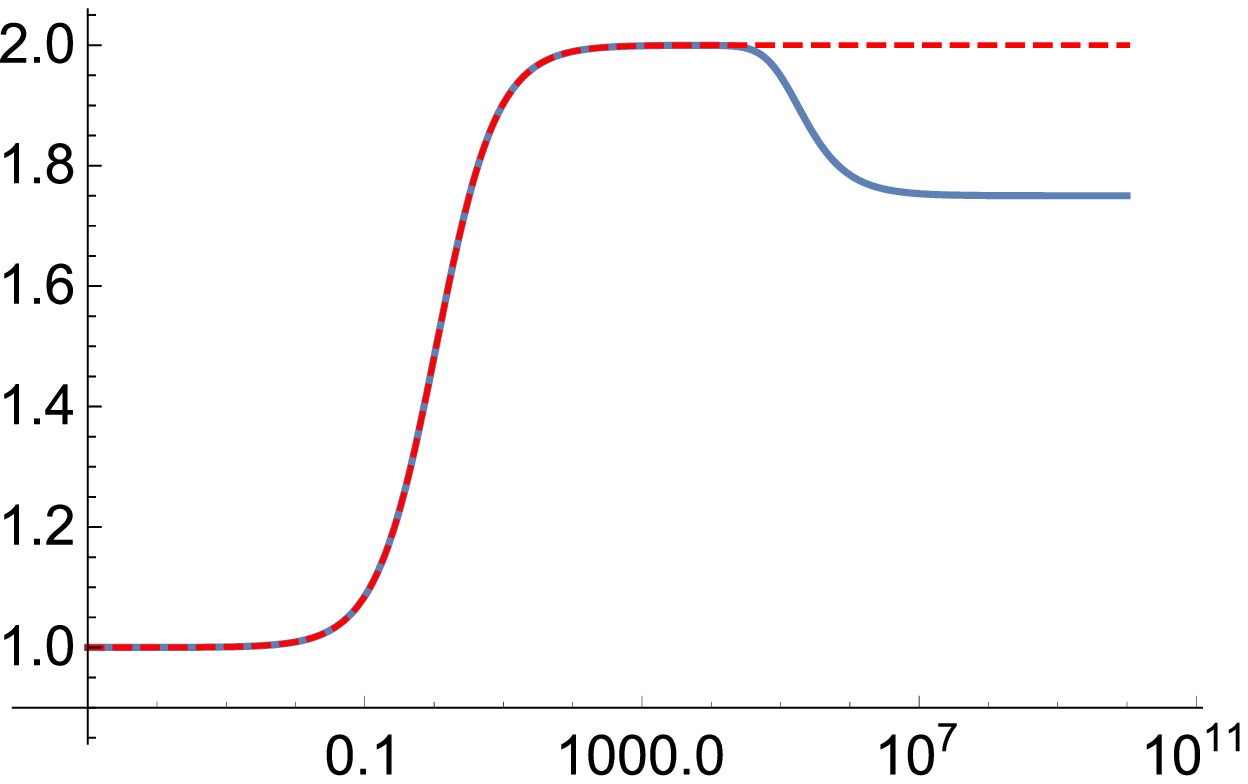}
\put(3,5){ $r$}
\put(-175,120){$r \, \psi'$}
\end{subfigure}
\caption{Plots of the logarithmic derivatives of the metric functions. The red, dashed line corresponds to the metric functions in the asymptotically AdS$_{4}$ solution (\ref{metric_sol_analytic}). The blue, straight curve was produced numerically with $\,(c_1,c_2)=(0,-10^{-8})\,$.
\label{fig.almostAdS}}
\end{center}
\vspace{-1mm}
\end{figure}

Leaving aside the asymptotic behaviour of the tensor field, we now proceed to characterise solutions lying at the boundary of the $\,(c_1,c_2)\,$ parameter space. The shaded region of solutions in Figure~\ref{fig.region} is delimited. The upper (red line) and lower (brown line) boundaries yield configurations that do not approach the DW$_{4}$ solution (\ref{D2-scaling}) but acquire non-relativistic behaviours in the UV. For instance, the (blue) circle approaches a Lifshitz spacetime with $\,z=2\,$ whereas the (green) square approaches a conformally Lifshitz spacetime with $\,(z,\theta)=(1.86,-0.705)\,$. Lastly, the (black) rhombus at the origin of the parameter space $\,(c_{1},c_{2})=(0,0)\,$ is special and produces the asymptotically AdS$_{4}$ solution with constant scalars in (\ref{charges_sol_analytic})-(\ref{metric_sol_analytic}). This is the only point in Figure~\ref{fig.region} satisfying $\,c_1 + 3 \, c_2=0\,$, or equivalently, setting to zero the irrelevant deformations in (\ref{eq.irrelevantIR}) for the dilaton $\,e^\phi\,$ in the universal hypermultiplet. Moving slightly away from this point into the shaded region modifies the UV behaviour of the solution making it flow to the DW$_{4}$. We show this behaviour in Figure~\ref{fig.almostAdS} where we have produced the plot by setting $\,(c_1,c_2)=(0,-10^{-8})\,$. One sees that the logarithmic derivatives of the metric functions coincide quite accurately with the ones dictated by the asymptotically AdS$_{4}$ solution in  (\ref{charges_sol_analytic})-(\ref{metric_sol_analytic}) (red, dashed line) up to a value of the radial coordinate beyond which the functions in our ansatz transition to that of the $\,\mathrm{DW}_4\,$ asymptotics \eqref{D2-scaling}.

\subsection{Non-relativistic UV asymptotics}

As previously mentioned, the solutions associated with the points at the boundary of the shaded region in Figure~\ref{fig.region} have a non-relativistic scaling in the UV.\footnote{An analytic solution of this type was found in the $\,\mathcal{N}=2\,$ model of \cite{Chimento:2015rra} with a prepotential $\,\mathcal{F}=-iX^{0}X^{1}\,$. However, unlike in our model, the U(1) factor of the gauge group therein was gauged by the (electric) \mbox{graviphoton}.} An example of this behaviour is given by the (blue) circle in that figure, for which the BPS solution asymptotes a scaling solution with broken Lorentz symmetry
\be
\label{eq.zis2scaling}
e^{2U} \sim r^2 \ , \qquad e^{2(\psi-U)} \sim r \ , \qquad \beta \sim 0 \ , \qquad b_0 \sim r \ ,
\ee
and constant scalars at large values of the radial coordinate. This corresponds to a non-relativistic metric of the Lifshitz type with dynamical exponent $z=2$. Along the boundary line that joins the (blue) circle and the (black) rhombus from above (red line), the scaling solution \eqref{eq.zis2scaling} receives some logarithmic corrections that we have not investigated in detail.

\begin{figure}[t!]
\begin{center}
\begin{subfigure}[b]{0.4\textwidth}
\includegraphics[width=\textwidth]{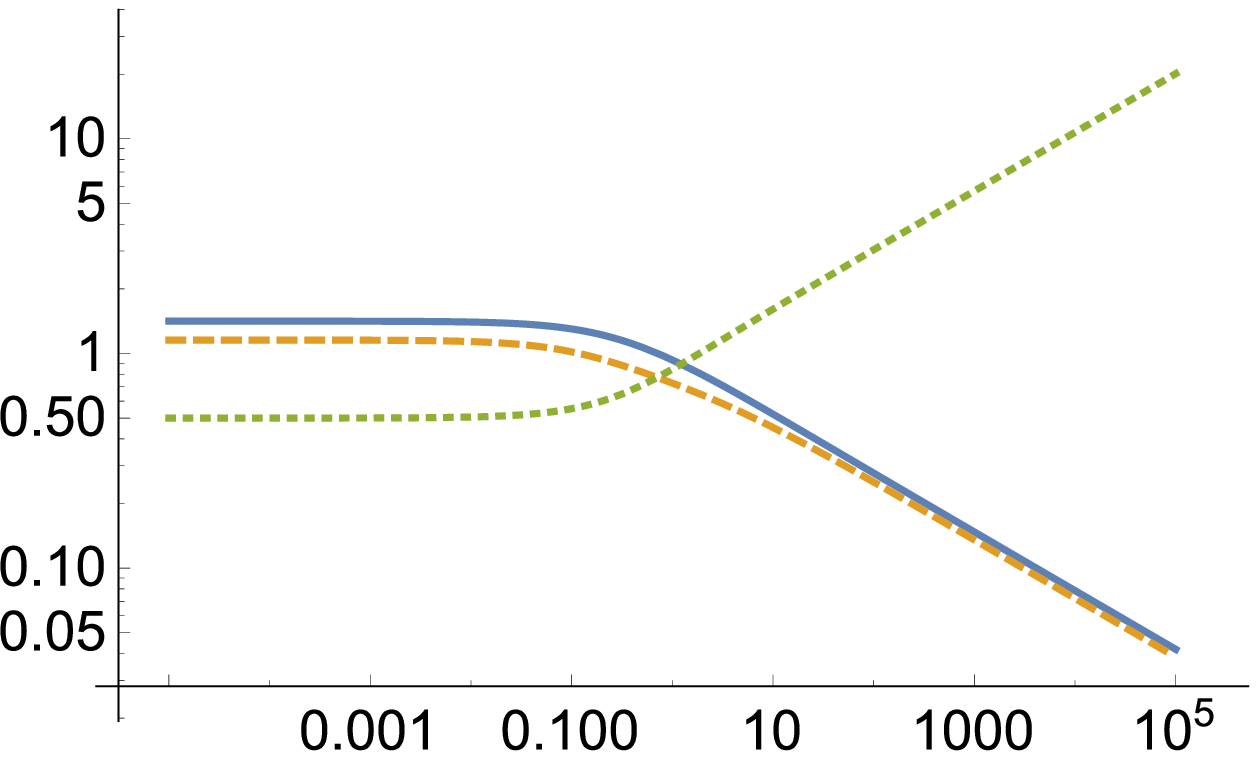}
\put(3,5){ $r$}
\put(-185,120){ {\color{NavyBlue}$e^\phi$}, {\color{Tan}$e^\varphi$}, {\color{LimeGreen}$-\chi$}}
\end{subfigure}\vspace{2mm}
~~~~~~
\begin{subfigure}[b]{0.4\textwidth}
\includegraphics[width=\textwidth]{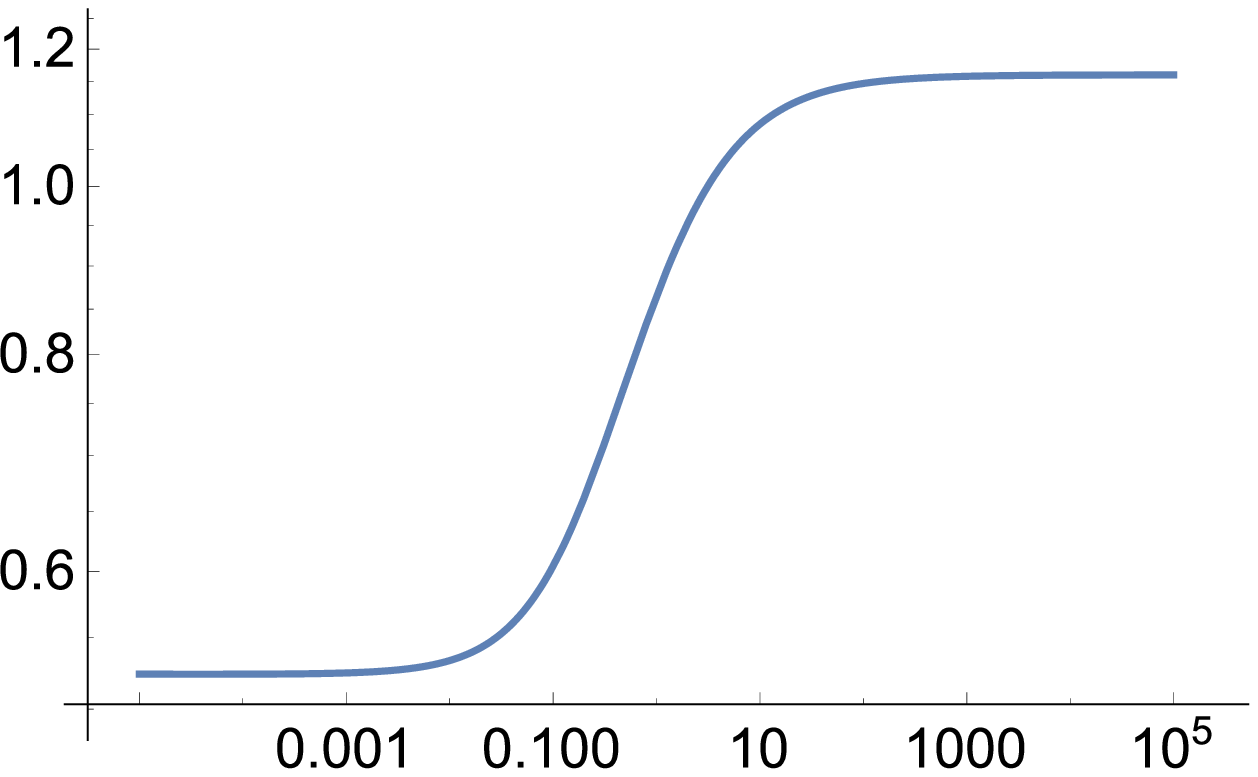}
\put(3,5){ $r$}
\put(-180,120){\small $-\beta$}
\end{subfigure}
\caption{Plots of the scalars $\,e^\phi\,$ (blue, straight line), $\,e^\varphi\,$ (brown, dashed line) and $\,-\chi\,$ (green, dotted line), as well as of the phase $\,-\beta$, as a function of the radial coordinate for a solution with $\,(c_1,c_2)=(1.138,-1.68)\,$.
\label{fig.newscaling}}
\end{center}
\end{figure}

A different non-relativistic scaling in the UV occurs for solutions associated with the points in the boundary line connecting the (blue) circle and the (black) rhombus in Figure~\ref{fig.region} from below (brown line). At large values of the radial coordinate, the solutions approach a behaviour of the form
\be
\begin{split}
e^{2U} & \sim r^{1.7268} \ , \qquad e^{2(\psi-U)} \sim r^{1.0484} \ , \qquad b_0 \sim r^{0.50197} \ , \\[2mm]
\chi & \sim r^{0.27325} \ , \qquad e^\phi \sim r^{-0.27325} \ , \qquad e^\varphi \sim r^{-0.27325} \ ,
\end{split}
\ee 
with $\,\beta\sim -1.1597\,$. A solution featuring this scaling in the UV is the one associated with the (green) square located at $\,(c_1,c_2)=(1.138,-1.68)\,$ in Figure~\ref{fig.region}, which we  present in Figure~\ref{fig.newscaling}. This solution can be written in the form of a non-relativistic metric conformal to a Lifshitz spacetime, characterised by a dynamical exponent $\,z=1.86\,$ and a hyperscaling violation parameter $\,\theta=-0.705\,$.

\section*{Acknowledgments}

We would like to thank Nikolay Bobev and Aristomenis Donos for conversations and Oscar Varela for collaboration in related work. The work of AG is partially supported by a Marina Solvay fellowship and by F.R.S.-FNRS through the conventions PDRT.1025.14 and IISN-4.4503.15. JT is supported by the Advanced ARC project ``Holography, Gauge Theories and Quantum Gravity" and by the Belgian Fonds National de la Recherche Scientifique FNRS (convention IISN 4.4503.15).

\appendix

\section{Equations of motion}
\label{App:EOMs}

The equations of motion can be found straightforwardly from (\ref{Lagrangian_model}). Let us start with the equation for $\,\mathcal{A}^0\,$ which takes the form
\be
d \! \left(  \cI_{0\Lambda} * \! H^\Lambda + \cR_{0\Lambda} \, H^\Lambda \right) = \frac{1}{2} \, g \, e^{4 \phi} * \! \left[ D a + \frac{1}{2} \left( \zeta \, D \tilde \zeta - \tilde \zeta \, D \zeta \right) \right]  \ ,
\ee
which can be seen to follow from \eqref{EOM_auxiliary} by taking an exterior derivative in the second one and using the first. Then, the equation of motion for $\,\mathcal{A}^0\,$ is redundant. On the other hand, the equation of motion for $\,\mathcal{A}^1\,$ reads
\be\begin{split}
\label{eq.A1vector}
d \! \left(  \cI_{1\Lambda} * \! \mathcal{H}^\Lambda + \cR_{1\Lambda} \, \mathcal{H}^\Lambda \right) & = \frac{3}{2} \, g \, e^{4 \phi} \left( \zeta^2 + \tilde \zeta^2 \right) * \! \left[ D a + \frac{1}{2} \left( \zeta \, D \tilde \zeta - \tilde \zeta \, D \zeta \right) \right] \\
	& \quad - \frac{3}{2} \, g \, e^{2\phi} \left( \tilde \zeta * \! D \zeta - \zeta * \! D \tilde \zeta \right) \ .
\end{split}
\ee
In the case when $\,\zeta=\tilde \zeta=0\,$, which is the relevant one in this work, it provides a first integration of motion since the right hand side in (\ref{eq.A1vector}) vanishes. 

We turn our attention now to the scalars. First let us consider $\,a\,$ in the universal hypermultiplet. Its equation of motion reads
\be
d \! \left[  e^{4 \phi} * \! \left( D a + \frac{1}{2} \left( \zeta \, D \tilde \zeta - \tilde \zeta \, D \zeta \right) \right)  \right] = 0 \ ,
\ee
which is a consequence of acting with $\,d\,$ on the right-hand side equation of the first equation in  \eqref{EOM_auxiliary}. Therefore, it is not an independent equation of motion. The scalars $\,\zeta\,$ and $\,\tilde \zeta\,$ satisfy the following equations
\be
\begin{split}
\label{eq.eomzetazetat}
\frac{1}{2} d \! \left[ e^{2 \phi} * \! D \zeta \right] & = \phantom{-}\frac{3}{2} \, g \, e^{2 \phi} \mathcal{A}^1 \wedge \! * D \tilde \zeta + \frac{1}{2} \, e^{4 \phi} \, D \tilde \zeta \wedge \! * \! \left[ D a + \frac{1}{2} \left( \zeta \, D \tilde \zeta - \tilde \zeta \, D \zeta \right) \right] + \partial_\zeta V_{g} * \! 1  \ , \\
\frac{1}{2} d \! \left[ e^{2 \phi} * \! D \tilde \zeta \right] & = - \frac{3}{2} \, g \, e^{2 \phi} \mathcal{A}^1 \wedge \! * D  \zeta - \frac{1}{2} \, e^{4 \phi} \, D  \zeta \wedge \! * \! \left[ D a + \frac{1}{2} \left( \zeta \, D \tilde \zeta - \tilde \zeta \, D \zeta \right) \right] + \partial_{\tilde \zeta} V_{g} * \! 1  \ ,
\end{split}
\ee
whereas the equation of motion for $\,\phi\,$ reads
\be
\begin{split}
\label{eq.phieom}
2 \, d \! * \! d \phi & =  e^{4 \phi} \left[ D a + \frac{1}{2} \left( \zeta \, D \tilde \zeta - \tilde \zeta \, D \zeta \right) \right] \wedge \! * \! \left[ D a + \frac{1}{2} \left( \zeta \, D \tilde \zeta - \tilde \zeta \, D \zeta \right) \right] \\
& \quad + \frac{1}{2} \, e^{2 \phi} \left[ D \zeta \wedge \! * D \zeta + D \tilde \zeta \wedge \! * D \tilde \zeta \right] + \partial_\phi V_{g} * \! 1 \ .
\end{split}
\ee
The scalars in the vector multiplet satisfy the equations of motion
\be
\label{eq.varphieom}
\frac{3}{2} \, d \! * \! d \varphi = \frac{3}{2} \, e^{2 \varphi} \, d \chi \wedge \! * d \chi - \frac{1}{2} \partial_\varphi \cI_{\Lambda \Sigma} \, \mathcal{H}^\Lambda \wedge \! * \mathcal{H}^\Sigma - \frac{1}{2}  \partial_\varphi \cR_{\Lambda \Sigma} \, \mathcal{H}^\Lambda \wedge  \mathcal{H}^\Sigma + \partial_\varphi V_{g} * \! 1\ ,
\ee
and
\be
\label{eq.chieom}
\frac{3}{2} \, d \! \left[ e^{2 \varphi} * \! d \chi \right] = - \frac{1}{2} \partial_\chi \cI_{\Lambda \Sigma} \, \mathcal{H}^\Lambda \wedge \! * \mathcal{H}^\Sigma - \frac{1}{2}  \partial_\chi \cR_{\Lambda \Sigma} \, \mathcal{H}^\Lambda \wedge  \mathcal{H}^\Sigma + \partial_\chi V_{g} * \! 1 \ .
\ee

\noindent Finally, the Einstein equations are given by
\be
 R_{\mu \nu} - \frac{1}{2} \, g_{\mu \nu} \, R = T_{\mu \nu}^\mt{scalars} + T_{\mu \nu}^\mt{vectors} \ ,
\ee
with
\be
\begin{split}
T_{\mu\nu}^\mt{vectors} & = - \cI_{ \Lambda \Sigma } \left[  \mathcal{H}^\Lambda_{\mu \rho} \, {\mathcal{H}^{\Sigma}_\nu}^{\rho}  -  \frac{1}{4} \, g_{\mu \nu}  \, \mathcal{H}^\Lambda_{ \rho \sigma} \, {\mathcal{H}^{\Sigma}}^{\rho \sigma}  \right] \ , \\
T_{\mu\nu}^\mt{scalars} & =  \frac{3}{2} \, \left( \partial_\mu \varphi \, \partial_\nu \varphi   - \frac{1}{2}  \, g_{\mu \nu}    \, \partial_\rho \varphi \, \partial^\rho \varphi  \right) +  \frac{3}{2} \, e^{2 \varphi} \, \left(   \partial_\mu \chi \, \partial_\nu \chi   - \frac{1}{2} \, g_{\mu \nu}   \, \partial_\rho \chi \, \partial^\rho \chi   \right)  \\
& \quad + 2 \, \left( \partial_\mu \phi \, \partial_\nu \phi - \frac{1}{2} \, g_{\mu \nu}  \, \partial_\rho \phi \, \partial^\rho \phi \right) + \frac{1}{2}\, e^{2 \phi} \, \left( D_\mu \zeta \, D_\nu \zeta - \frac{1}{2} \, g_{\mu\nu} \, D_\rho \zeta \, D^\rho \zeta \right)   \\
& \quad + \frac{1}{2}\, e^{2 \phi} \, \left( D_\mu \tilde \zeta \, D_\nu \tilde \zeta - \frac{1}{2} \, g_{\mu\nu} \, D_\rho \tilde \zeta \, D^\rho \tilde \zeta   \right) + \frac{1}{2} \, e^{4 \phi}  \left( \xi_\mu \, \xi_\nu - \frac{1}{2} \, g_{\mu \nu}  \, \xi_\rho \, \xi^\rho \right) \\
& \quad -  g_{\mu \nu}  \, V_{g}  \ ,
\end{split}
\ee
and where, for presentational convenience, we have introduced the quantity
\be
\xi_\mu \equiv D_\mu a + \frac{1}{2} \, \left( \zeta \, D_\mu \tilde \zeta - \tilde \zeta \, D_\mu \zeta \right) \ .
\ee

\section{UV expansion around DW$_{4}$}
\label{App:UV_expansion}

In this appendix we provide the UV asymptotic expansion of the solution to the equations of motion presented in appendix~\ref{App:EOMs} around the four-dimensional DW$_{4}$ description of the \mbox{D2-brane}.

As described at the end of section~\ref{sec:ansatz}, the set of equations of motion to be solved consists of five second order differential equations (for $U$, $\psi$, $\chi$, $\varphi$ and $\phi$), two first order ones \mbox{(for $b_0$ and $\alpha_t^-$)} and a first order constraint, giving a total of eleven constants of integration. However, one must also integrate (\ref{Atprim_eqs}) which, subject to the constraints in (\ref{EOM_A0tilde}), gives a new constant of integration identified with the constant value of the field $\,\mathcal{A}_{t}{}^{1}\,$. Altogether, we denote the twelve constants of integration with capital latin characters
\be
\label{constants1}
\{T_0,\ T_1,\ T_2,\ S_0,\ F_1,\ F_2,\ F_3,\ C_1,\ C_2,\ B_1,\ A_1 ,\ A_2 \} \ ,
\ee
and the UV expansions of the metric, scalar, vector and tensor fields depend on them as well as on the parameters of the theory
\be
\label{parameters1}
\{ p^0 , \ p^1 , \ e_1 , \ \kappa , \ m , \ g \} \ ,
\ee
which we decide to keep unfixed in this exposition. Recall that $\,e_{0}=\frac{g}{m} \, p^{0}\,$ by virtue of (\ref{EOM_A0tilde}). In our numerical integration all the constants in \eqref{constants1} will depend parametrically on the two IR parameters $\,(c_1,c_2)\,$ of Figure~\ref{fig.region}, thus providing non-trivial relations amongst them.

The metric functions $\,U(r)\,$ and $\,\psi(r)\,$ have a UV expansion around the DW$_{4}$ solution given by
\be
\label{metric_UV_expansion}
\begin{split}
e^{2U} & = T_0 \, r^{7/4} \Bigg( 1 + \frac{T_1}{r} + \frac{\frac{2\, \kappa}{S_0\, T_0} + \frac{m^2 T_0^6}{256\, g^{14}} - \frac{C_1^2 \, T_0^2}{4\, g^4} - \frac{40\, B_1 \, g^8}{3\, S_0^2 \, T_0^4} }{r^{3/2}} \\
& \qquad \qquad \,\,\,\,\,\,\,\,\,+ \frac{T_1^2 - 4 \, F_1\, T_1 -14\, F_1^2}{6\, r^{2}} + \frac{T_2 }{r^{5/2}} + {\cal O}(r^{-3}) \Bigg) \ , \\[3mm]
e^{2(\psi-U)} & = S_0 \, r^{7/4} \Bigg( 1 + \frac{T_1}{r} + \frac{\frac{2\, \kappa}{3\, S_0\, T_0} + \frac{m^2 T_0^6}{256\, g^{14}} - \frac{C_1^2 \, T_0^2}{4\, g^4} - \frac{8\, B_1 \, g^8}{S_0^2 \, T_0^4} }{r^{3/2}}  \\
& \qquad \qquad \,\,\,\,\,\,\,\,\, + \frac{T_1^2 - 4 \, F_1\, T_1 -14\, F_1^2}{6\, r^{2}}  + {\cal O}(r^{-5/2}) \Bigg)\ .
\end{split}
\ee
The constants $T_0$ and $S_0$ are related to a global rescaling of the time coordinate and the symmetry in \eqref{rescaling_sym}. In the former (and following) expansions the omitted higher order terms are algebraically determined in terms of  the constants in \eqref{constants1} and the parameters in \eqref{parameters1}. More concretely, specific powers of the charges appear. 

The scalars $\,e^{\varphi(r)}\,$ and $\,\chi(r)\,$ in the vector multiplet have a near UV expansion of the form
\be
\begin{split}
e^\varphi & = \frac{T_0}{2\, g^2\, r^{1/4}} \Bigg( 1 - \frac{T_1 + 4 \, F_1}{3\, r} - \frac{\frac{2\, \kappa}{3\, S_0\, T_0} + \frac{m^2 T_0^6}{256\, g^{14}} - \frac{C_1^2 \, T_0^2}{6\, g^4} - \frac{56\, B_1 \, g^8}{9\, S_0^2 \, T_0^4} - \frac{4\, F_2}{3}}{r^{3/2}} \\[1mm]
& \qquad \qquad \qquad + \frac{T_1^2 + 4 \, F_1\, T_1 -6\, F_1^2}{6\, r^{2}}  + {\cal O}(r^{-3})  \Bigg) \ , \\[3mm]
\chi & = \frac{C_1}{r^{1/2}} + \frac{C_2}{r^{3/2}} \\[1mm]
& \,\,\,\,\,\,\, + \left[ C_1 \left( \frac{4\, \kappa}{9\, S_0\, T_0} + \frac{m^2 T_0^6}{384\, g^{14}} + \frac{80\, B_1^2 \, g^8}{27\, S_0^2 \, T_0^4} + \frac{8\, F_2}{9} \right) + \frac{5\, C_1^3\, T_0^2}{36\, g^4} - \frac{m\, C_1^2 \, T_0^4}{48\, g^9} - \frac{32\, B_1\, p^1\, g^7}{3\, S_0^2 \, T_0^4}   \right] \frac{1}{r^2} \\[1mm]
& \,\,\,\,\,\,\, + {\cal O}(r^{-5/2})\ , 
\end{split}
\ee
whereas the UV expansion of the non-trivial dilaton $\,e^{\phi(r)}\,$ in the universal hypermultiplet reads
\be
e^\phi = \frac{T_0}{2\, g^2\, r^{1/4}} \left( 1 + \frac{F_1}{r} + \frac{F_2}{r^{3/2}} + \frac{5\, F_1^2}{2\, r^2} + \frac{F_3}{r^{5/2}} + {\cal O}(r^{-3})  \right) \ .
\ee

From the above expansions one can extract the UV behaviour of the vector and tensor fields using (\ref{Atprim_eqs}) in combination with (\ref{EOM_A0tilde}). The vector fields are given by
\be
\begin{split}
\mathcal{A}_t{}^0 & = - \frac{8\, B_1 \, g^7}{S_0\, T_0^3} r^{1/2} +  \frac{A_2}{r^{1/2}} + {\cal O}(r^{-1}) \ , \\[2mm]
\tilde{\mathcal{A}}_{t \, 0} & = - \frac{m\, B_1 \, T_0^3}{16\, S_0\, g^6} \frac{1}{r}  + {\cal O}(r^{-2}) \ ,\\[2mm]
\mathcal{A}_t{}^1 & = A_1 + \frac{e_1\, T_0}{6\, S_0\, g^2} \frac{1}{r}  + {\cal O}(r^{-3/2}) \ ,
\end{split}
\ee
whereas the tensor field reads
\be
\label{b0_UV_expansion}
b_0 =  B_1 \, r^{1/2} - \frac{2\, p^0}{m} + \left( \frac{A_2 \, S_0 \,  T_0^3}{8\, g^7} - 4\, B_1\, F_1 \right) \frac{1}{r^{1/2}} + {\cal O}(r^{-1}) \ .
\ee
Note that, despite $\,\mathcal{A}_t{}^0\,$ and $\,b_0\,$ having a positive power of the radial coordinate $\,r\,$ governed by the integration constant $\,B_1\,$, their norm remains finite in the UV due to the higher powers of $\,r\,$ that appear in the metric functions.

For the sake of completeness, we present also the UV expansion of the angle $\,\beta\,$ in \eqref{W_func}, which can be obtained algebraically
\be
\beta = \left( \frac{m \, T_0^3}{56\, g^7}+ \frac{4\, B_1\, g^3}{21\, p^0\, S_0\, T_0^2\, \kappa} + \frac{9\, C_1\, T_0}{14\, g^2}  \right) \frac{1}{r^{3/4}} + {\cal O}(r^{-7/4}) \ .
\ee

Once we have the UV expansion of the second order differential equations in appendix~\ref{App:EOMs} around the DW$_4$ solution, we now move to analyse the BPS equations in \eqref{BPS-eqs}. For these equations to hold the quantisation condition \eqref{quant_cond} must be imposed, thus fixing
\be
p^1= \frac{1}{3\, g\, \kappa} \ . 
\ee
Since the set of BPS equations in \eqref{BPS-eqs} consists of six first order differential equations (plus an algebraic equation for $\beta$), we expect the system to be determined by six constants of integration.\footnote{Together with the additional constant of integration $A_1$ obtained from the integration of \eqref{Atprim_eqs}.} Therefore, the BPS equations must provide five relations between the integration constants in \eqref{constants1}. This is indeed the case:
\be
\label{BPS-constraints_coefficients}
\begin{split}
B_1 & =\frac{3\, S_0\, T_0^3\, \left( 8\, C_1\, g^5 + m\, T_0^2 \right)}{128\, g^{11}} \ , \\[2mm]
F_2 & =\frac{3\, m^2 \, T_0^6 + 48\, C_1\, T_0^4\, m\, g^5 - 64 \,C_1^2\, T_0^2\, g^{10}}{1024\, g^{14}} \ , \\[2mm]
C_2 & = \frac{-16\, g^6\, e_1 - 24\, A_2\, g^4\, S_0\, T_0 - 24\, C_1\, g^5\, S_0\, T_1\, T_0 + m\, S_0 \, \left( 16 \, F_1 + T_1 \right)\,T_0^3}{48\, g^5\, S_0\, T_0} \ ,
\end{split}
\ee
and $\,(T_2,F_3)\,$ are similarly expressed in terms of the remaining six constants of integration $\{T_0,\ S_0,\ C_1,\ T_1,\ F_1,\ A_2\}$. The expressions for the latter are lengthy and not very enlightening, so we are not presenting them here.

An analysis of the numerics shows that $\,C_1<0\,$ in the solutions presented in the main text. For sufficiently large values of $\,|C_1|\,$ the constant $\,B_1\,$ in (\ref{BPS-constraints_coefficients}) becomes negative, whereas for small values it becomes positive. Since $\,B_1\,$ enters the expansion of the tensor field in (\ref{b0_UV_expansion}), one has a limiting case where
\be
\label{B1_cond}
B_1= 0 \qquad \Rightarrow \qquad C_1 = - \frac{m \, T_0^2}{8 \, g^5} 
\hspace{5mm} \textrm{ and } \hspace{5mm}
b_{0}(r)=- \frac{2\, p^0}{m} + {\cal O}(r^{-1/2})
\ .
\ee
Setting $\,T_0=1\,$ without loss of generality by a rescaling of the time coordinate, the condition (\ref{B1_cond}) determines a curve in parameter space, $\,C_1(c_1,c_2)=-m/(8 \, g^5)\,$, corresponding to the (grey) dashed line in Figure~\ref{fig.region}. The BPS flows in this curve have $\,\left. b_{0}(r) \right|_{r \rightarrow \infty} = - 2\, p^0/m\,$.

\bibliography{references}

\end{document}